\documentclass[preprint,12pt,a4paper]{elsarticle}

\makeatletter
\def\ps@pprintTitle{%
  \let\@oddhead\@empty
  \let\@evenhead\@empty
  \def\@oddfoot{\reset@font\hfil\thepage\hfil}
  \let\@evenfoot\@oddfoot
}
\makeatother

\usepackage{lineno,hyperref}
\modulolinenumbers[5]

\begin{document}


\begin{frontmatter}

\title{On the analogy between the bias flow aperture theory and the Coriolis
flowmeter "bubble theory"}

\author{Nils T. Basse\fnref{myfootnote}}
\address{Elsas v\"ag 23, 423 38 Torslanda, Sweden \\ \vspace{1cm} \today}
\fntext[myfootnote]{nils.basse@npb.dk}

\begin{abstract}
Two different physical phenomena, described by the bias flow aperture theory and the Coriolis flowmeter "bubble theory", are compared. The bubble theory is simplified and analogies with the bias flow aperture theory are appraised.
\end{abstract}


\end{frontmatter}


\section{Introduction}

In this paper, two phenomena which originate in different fields are treated:

The first phenomenon is acoustics of a bias flow aperture, where vortices are generated at the aperture edge. These vortices can (i) block the aperture and (ii) absorb acoustic energy. The linear theory was presented in \cite{howe_a,howe_b} and (small) corrections due to nonlinearity were treated in \cite{luong_a}. The findings can be applied to e.g. sound attenuation in wind tunnel guiding vanes.

The second phenomenon is the reaction force on an oscillating fluid-filled container due to entrained particles \cite{hemp_a}. This linear theory was motivated by the need to model two-phase flow in Coriolis flowmeters and is known as the "bubble theory". The bubble theory has been used to model both (i) measurement errors \cite{basse_a} and (ii) damping \cite{basse_b} experienced by Coriolis flowmetering of two-phase flow.

Theories for the two phenomena have been independently derived, the bubble theory about 25 years after the bias flow aperture theory. There are similarities between the two theories, and in this paper it will be studied how far this analogy can be taken.

In earlier work, two other physical phenomena have been identified which can be analysed by use of the bubble theory:
\begin{itemize}
  \item Two-phase flow damping in steam generators \cite{basse_b}
  \item Sound propagation in an aerosol \cite{basse_c}
\end{itemize}

The paper is organised as follows: In Section \ref{sec:bfa}, the bias flow aperture theory is briefly summarized. This is followed by a similar overview of the bubble theory in Section \ref{sec:bt}. First observed similarities between the two theories are introduced in Section \ref{sec:anal} along with simplifications of the bubble theory. The physical understanding which results is discussed in Section \ref{sec:disc}. Finally, conclusions are placed in Section \ref{sec:conc}.

\section{The bias flow aperture theory}
\label{sec:bfa}

\subsection{Linear theory}

\begin{figure}
\includegraphics[width=12cm]{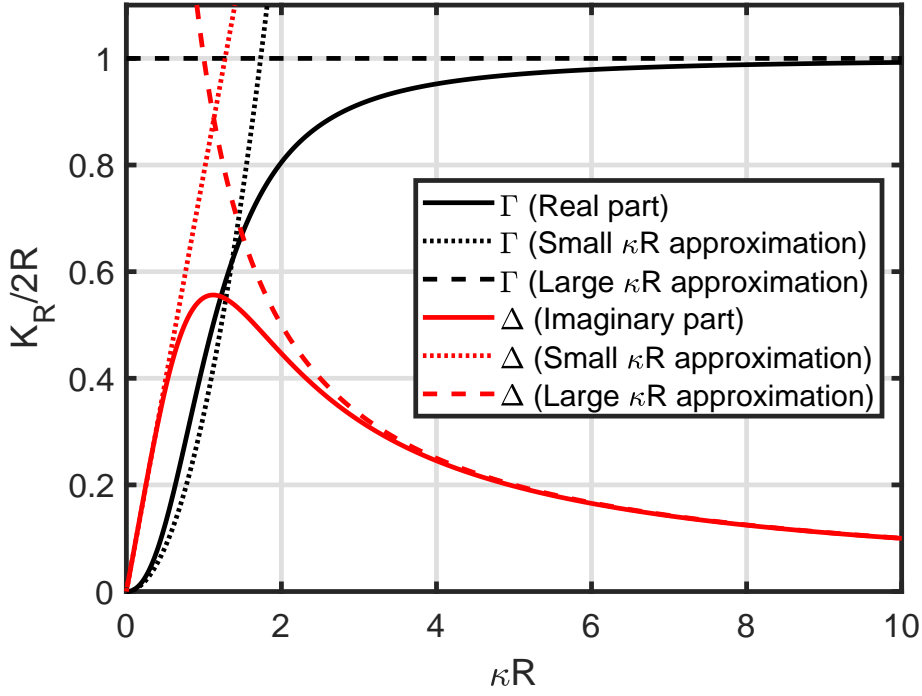}
\caption{Real and imaginary part of the normalised Rayleigh conductivity.}
\label{fig:Howe_1979}
\end{figure}

The bias flow aperture theory treats single-phase, low Mach number (incompressible) flow through a circular aperture in a rigid, thin plate \cite{howe_a,howe_b}. High Reynolds number ($Re$) flow is considered, so viscosity is only important at the rim of the aperture. Flow through the aperture creates a jet and vortex shedding from the aperture rim. The vortices lead to acoustic damping, since they absorb acoustic energy and are swept downstream by the jet. The vortices also lead to partial blockage of the flow; both damping and blockage can be characterised using the Rayleigh conductivity $K_R$ of aperture:

\begin{equation}
\label{eq:Rayleigh_def}
K_R=\frac{-\rm{i} \omega \rho_0 Q}{(p_+ - p_-)},
\end{equation}

\noindent where $\omega$ is the harmonic variation of the pressure difference between the high- ($p_+$) and low- ($p_-$) pressure sides of the aperture, $\rho_0$ is the mean density and $Q$ is the volume flux. As is the case for the pressure difference, the volume flux will also have a time variation ${\rm e}^{-{\rm i} \omega t}$, where $t$ is time. The mean jet velocity in the plane of the aperture is $U$ and the aperture radius is $R$. The Rayleigh conductivity can be written as a complex number:

\begin{equation}
\label{eq:howe_one}
 K_R = 2R \left( \Gamma - \rm{i} \Delta \right),
\end{equation}

\noindent where:

\begin{equation}
\label{eq:howe_two}
\Gamma - \rm{i} \Delta = 1+\frac{(\pi/2){\rm I}_1 (\kappa R){\rm e}^{- \kappa R} - {\rm i} {\rm K}_1 (\kappa R) \sinh (\kappa R)}{\kappa R [(\pi/2) {\rm I}_1 (\kappa R) {\rm e}^{-\kappa R}+{\rm i} {\rm K}_1 (\kappa R)\cosh(\kappa R)]},
\end{equation}

\noindent where ${\rm I}_1$ and ${\rm K}_1$ are modified Bessel functions and $\kappa \equiv \omega/U>0$, see Fig. \ref{fig:Howe_1979}. The Strouhal number is

\begin{equation}
\label{eq:Strouhal}
\kappa R = \omega R/U
\end{equation}

Asymptotic approximations of the normalised Rayleigh conductivity $K_R/2R$ are also shown in Fig. \ref{fig:Howe_1979}.

Small Strouhal number approximations (real part: Applicable for $\kappa R \leq 0.04$, imaginary part: Applicable for $\kappa R < 0.5$):

\begin{eqnarray}
  {\rm Re} (K_R/2R) = \Gamma & \approx & \frac{1}{3} (\kappa R)^2 \\
  {\rm Im} (K_R/2R) = -\Delta & \approx &  - \frac{ \pi}{4} \kappa R
\end{eqnarray}

Large Strouhal number approximations (applicable for $\kappa R > 5$):

\begin{eqnarray}
  {\rm Re} (K_R/2R) = \Gamma & \approx & 1 \\
  {\rm Im} (K_R/2R) = -\Delta & \approx & -\frac{1}{\kappa R}
\end{eqnarray}

\subsection{Nonlinear modelling and approximations}

\subsubsection{Thin wall}

It has been found that both linear and nonlinear regimes for bias aperture flow can be approximated by \cite{luong_a}:

\begin{equation}
\label{eq:Luong_thin}
K_R=2R \left( \frac{\omega R/U}{\omega R/U + 2\rm{i}/\pi \sigma^2} \right),
\end{equation}

\noindent where the contraction ratio of the jet $\sigma \approx 0.75$. The contraction ratio is the ratio of the jet area $\mathcal{A}$ at the {\it vena contracta} to the aperture area:

\begin{equation}
\label{eq:contraction_ratio}
\sigma=\frac{\mathcal{A}}{\pi R^2}
\end{equation}

The minimum theoretical value of $\sigma$ is $1/2$ \cite{howe_a}. Eq. (\ref{eq:Luong_thin}) can also be written:

\begin{equation}
\label{eq:reform_Luong_thin}
K_R/2R= \frac{\omega R/U}{\omega R/U + 2\rm{i}/\pi \sigma^2}
\end{equation}

\subsubsection{Wall of finite thickness}

In case the aperture wall has a finite thickness, the Rayleigh conductivity can be expressed as:

\begin{equation}
\label{eq:Luong_thick}
K_R=\frac{K_0 (\omega \ell /U)}{(\omega \ell / U)+{\rm i}/\sigma^2},
\end{equation}

\noindent where $K_0=\frac{\pi R^2}{\ell}$, $\ell = \pi R/2 + \ell_w$, where $\ell_w$ is the wall thickness. Eq. (\ref{eq:Luong_thick}) can be modified to an equation for the normalised Rayleigh conductivity:

\begin{equation}
\label{eq:reform_Luong_thick}
K_R/2R=\left( \frac{\pi R}{2 \ell} \right) \left[ \frac{\omega \ell / U}{\omega \ell / U + {\rm i}/\sigma^2} \right]
\end{equation}

\section{The Coriolis flowmeter bubble theory}
\label{sec:bt}

The Coriolis flowmeter bubble theory was first presented in \cite{hemp_a}. It is a linear theory for an incompressible, low Reynolds number flow. The force on a fluid-filled, oscillating container due to entrained particles is calculated. The particles can either be solid or consist of a fluid. The motion of the container leads to decoupled motion of the fluid and the particles, which leads to both (i) measurement errors and (ii) damping of Coriolis flowmeters. These effects have been studied in \cite{basse_a} and \cite{basse_b}, respectively. The entrained particles mean that a two-phase flow is considered by the theory. The force on the container is given by:

\begin{equation}
F_{f,z}=(\rho_f-\rho_p) V_p a_c F,
\label{eq:force_fluid}
\end{equation}

\noindent where $\rho_f$ is the fluid (f) density, $\rho_p$ is the particle (p) density, $V_p$ is the particle volume, $a_c$ is the container acceleration, $z$ is the acceleration direction and $F$ is the reaction force coefficient:

\begin{equation}
\label{eq:reaction_force}
F=1+\frac{4(1-\tau)}{4\tau-(9{\rm i} G/\beta^2)}
\end{equation}

\begin{figure}
\centering
\includegraphics[width=6cm]{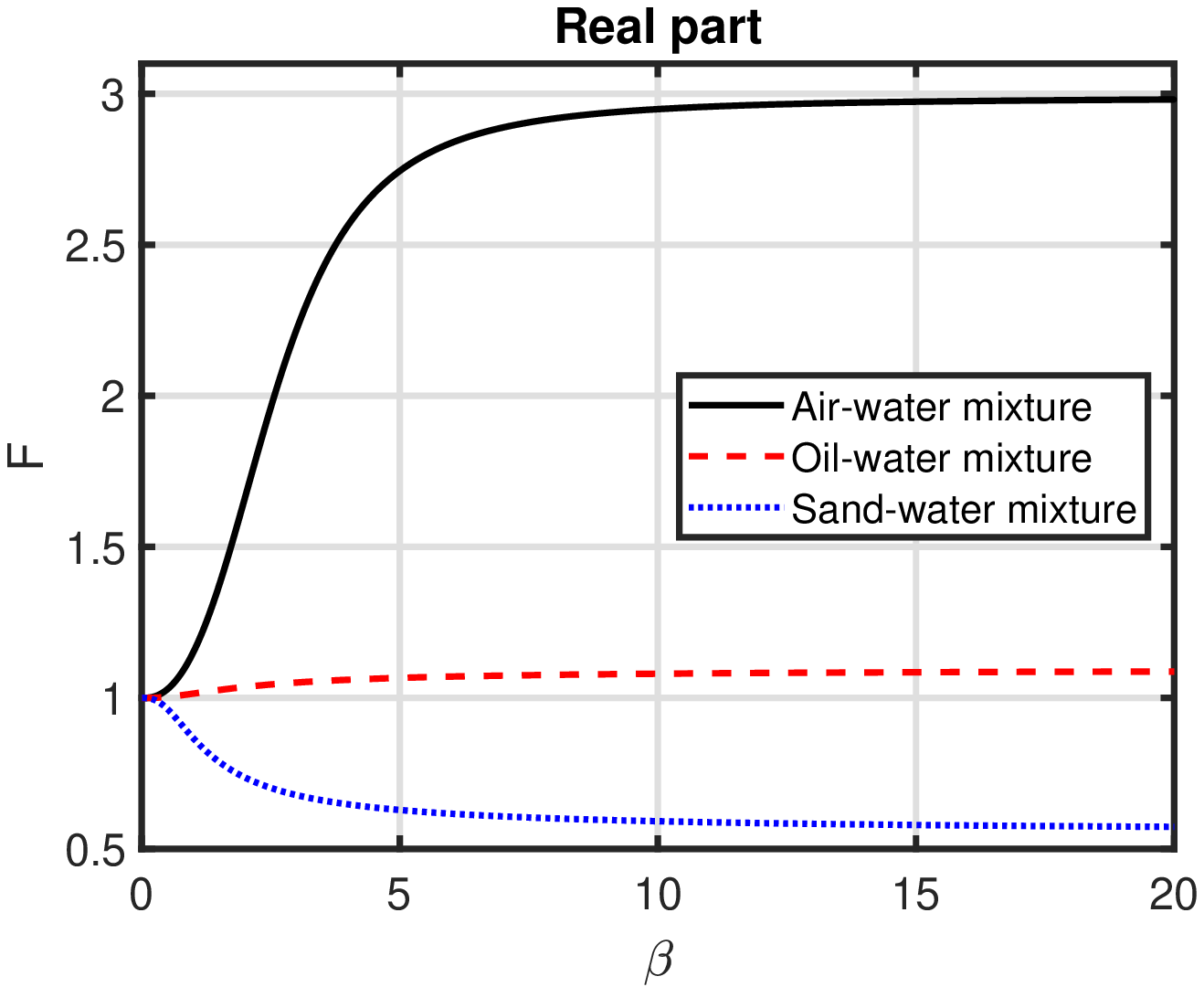}
\includegraphics[width=6cm]{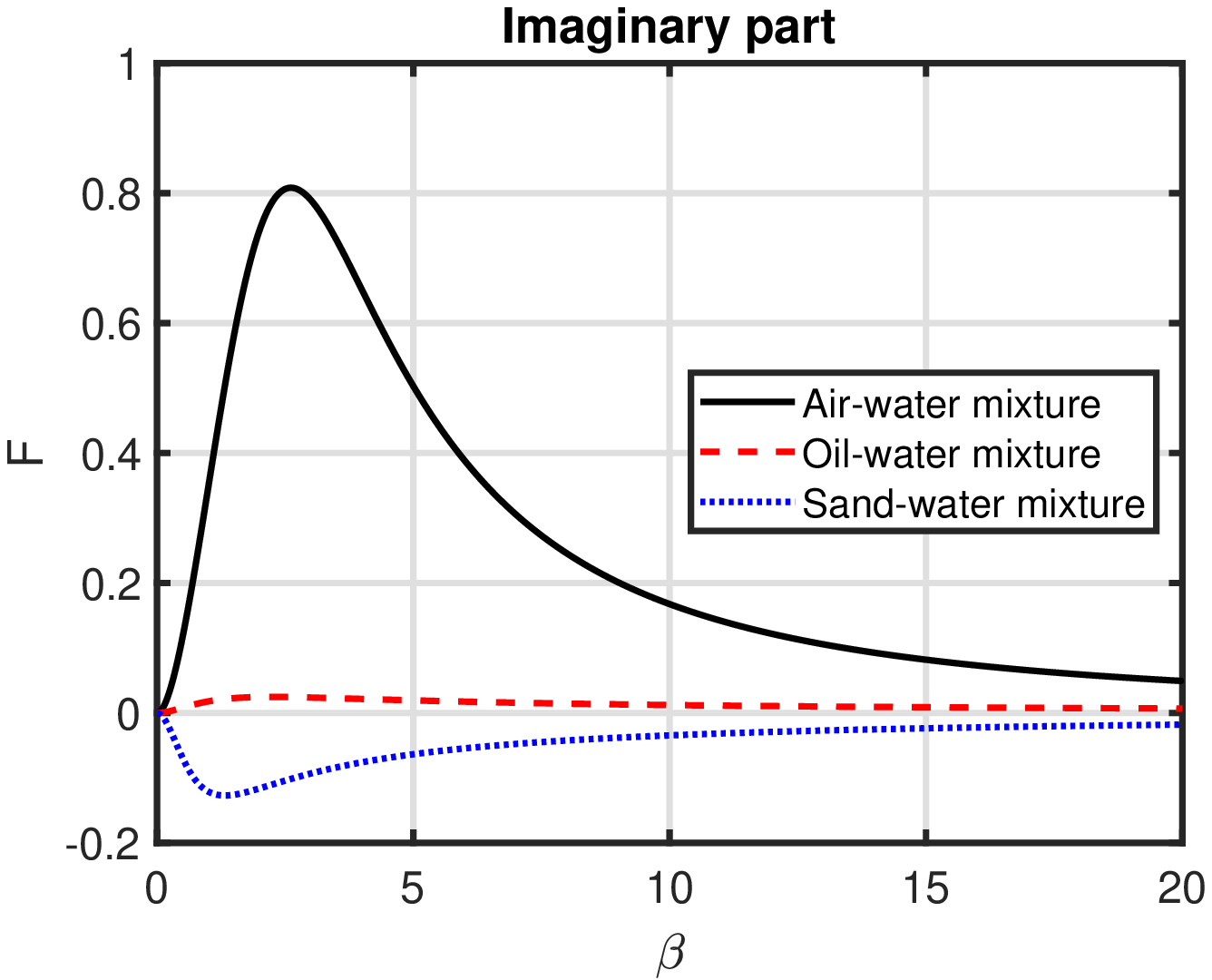}
\caption{Reaction force coefficients for three mixtures, left: Real part, right: Imaginary part.}
\label{fig:F_three_mixtures}
\end{figure}

The real part of $F$ is a virtual mass loss and the imaginary part of $F$ represents damping which acts against the vibrating force. This is exemplified for three mixtures in Fig. \ref{fig:F_three_mixtures}. More details on the material properties used for the mixtures are available in \cite{basse_a}.

The density ratio is

\begin{equation}
\label{eq:density_ratio}
\tau=\frac{\rho_p}{\rho_f}
\end{equation}

The Stokes number is

\begin{equation}
\label{eq:Stokes}
\beta=\frac{a}{\delta} = a \sqrt{\frac{\omega \rho_f}{2 \mu_f}},
\end{equation}

\noindent where $a$ is the particle radius, $\omega$ is the oscillation frequency of the container and $\mu_f$ is the dynamic viscosity of the fluid.

The quantities below are defined in \cite{yang_a}:

\begin{equation}
G = 1+\lambda+\frac{\lambda^2}{9}-\frac{(1+\lambda)^2f(\lambda)}{\kappa[\lambda^3 - \lambda^2 \tanh \lambda -2 f(\lambda)]+(\lambda+3)f(\lambda)},
\label{eq:big_g}
\end{equation}

\noindent where

\begin{equation}
\label{eq:lambda}
\lambda=(1+{\rm i})\beta
\end{equation}

\noindent and

\begin{equation}
\label{eq:small_f}
f(\lambda) = \lambda^2 \tanh \lambda - 3 \lambda + 3 \tanh \lambda
\end{equation}

The viscosity ratio is

\begin{equation}
\label{eq:visc_ratio}
\kappa=\frac{\mu_p}{\mu_f}
\end{equation}

$G$ is "proportional to the drag force on a spherical particle undergoing harmonic motion in a surrounding (stagnant) liquid" \cite{hemp_a} (fluid):

\begin{equation}
 F_D=-u_p (6 \pi \mu_f a G),
 \end{equation}

\noindent where $u_p$ is the particle velocity. Note that if $G=1$, the drag force reduces to the Stokes drag:

\begin{equation}
\label{eq:drag_Stokes}
F_{D, {\rm Stokes}} = -u_p (6 \pi \mu_f a)
\end{equation}

\section{Analogies}
\label{sec:anal}

\subsection{Initial observed similarities}

\begin{figure}
\centering
\includegraphics[width=6cm]{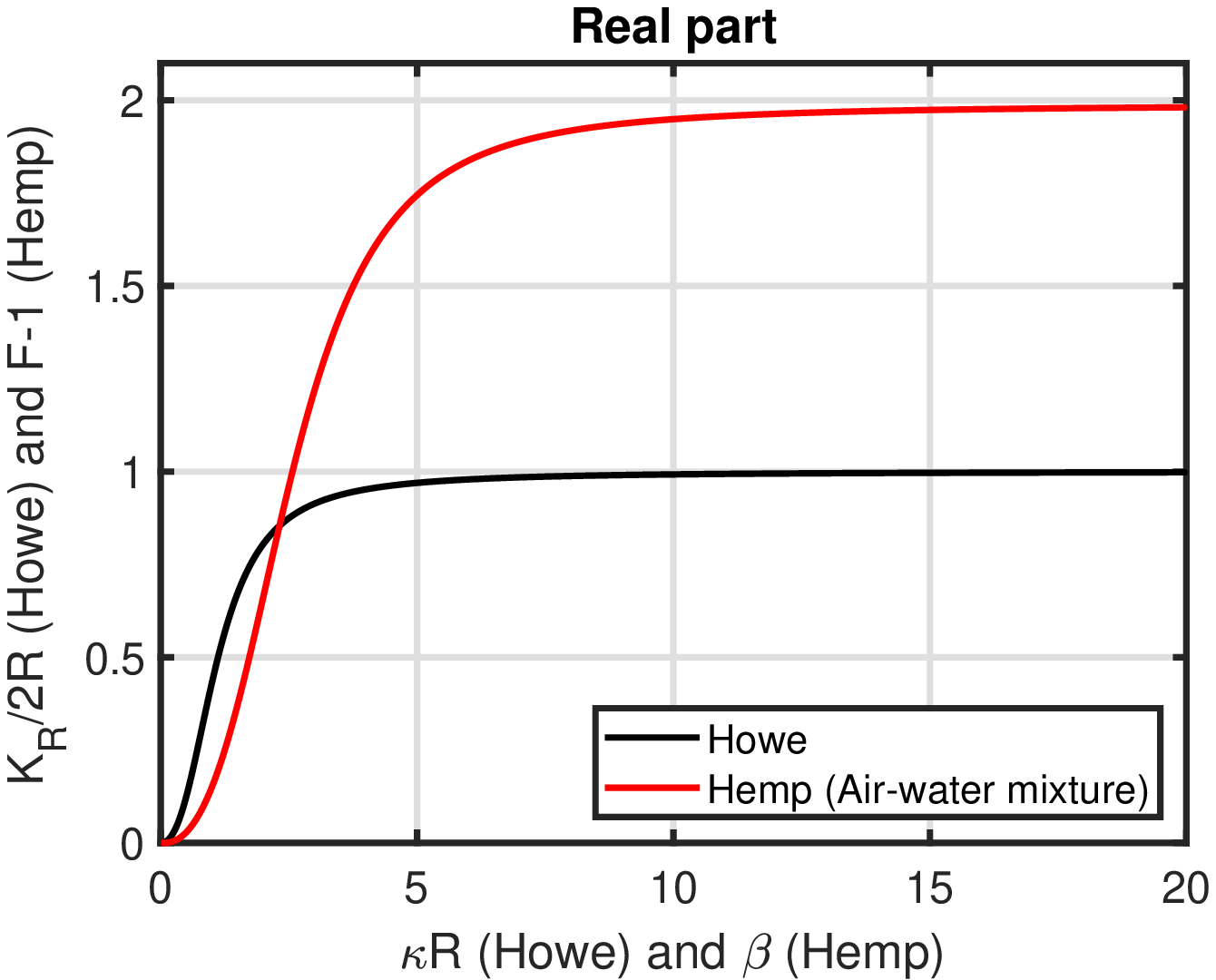}
\includegraphics[width=6cm]{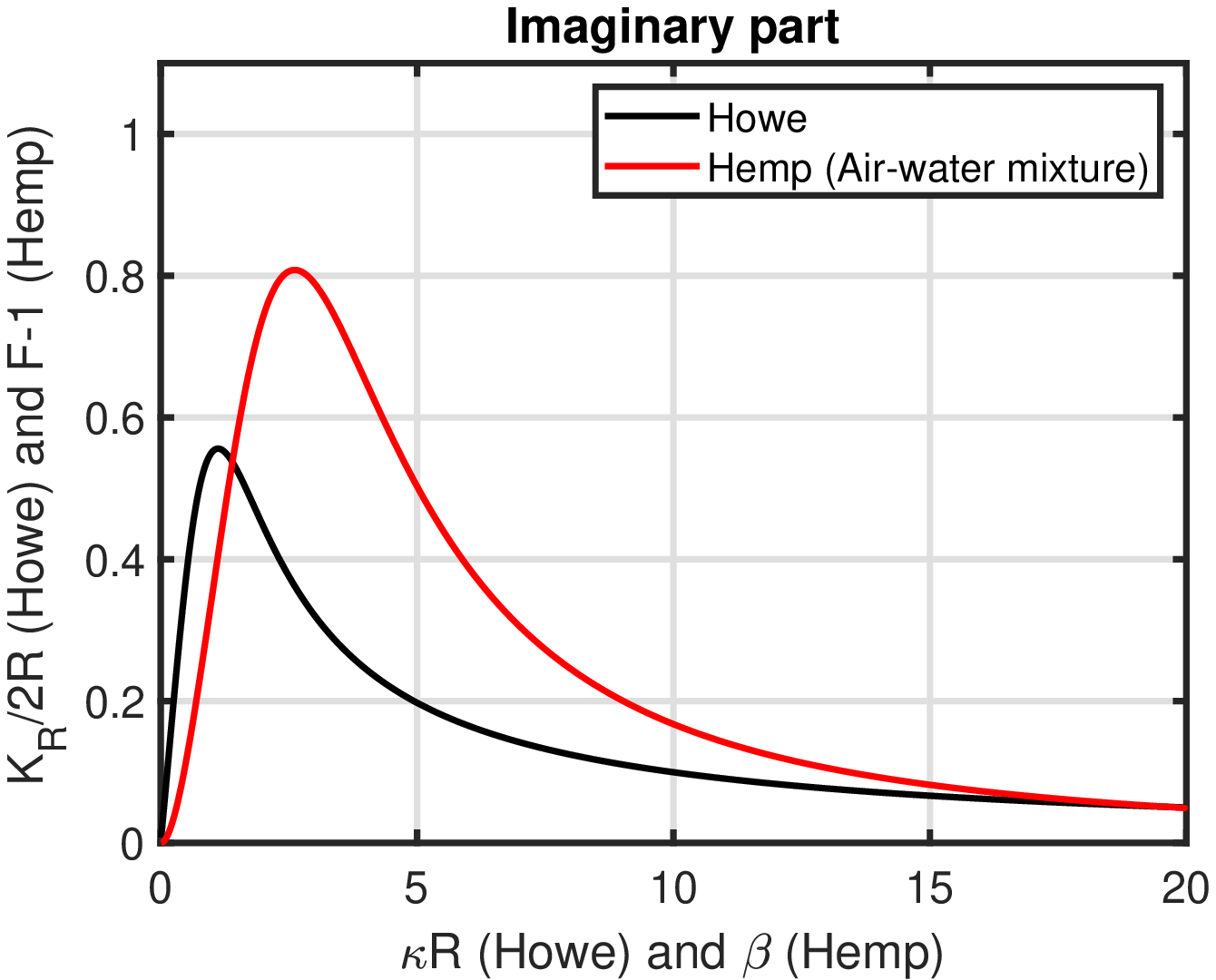}
\caption{Normalised Rayleigh conductivity and $F-1$ for an air-water mixture, left: Real part, right: Imaginary part.}
\label{fig:howe_hemp}
\end{figure}

The similarity between the normalised Rayleigh conductivity and the reaction force coefficient (minus 1) was most apparent for the case where the bubble theory is used for air bubbles in a water-filled container, see Fig. \ref{fig:howe_hemp}. The curve shapes are similar, both when comparing real and imaginary parts.

The physical phenomena are different, but have common features such as a characteristic angular frequency $\omega$, which is either the variation of the pressure difference across the aperture or the container oscillation. A characteristic size for both phenomena is also apparent, the aperture radius $R$ and the particle radius $a$. These observations are combined in that the Strouhal (Stokes) number is proportional to $\omega R$ ($a \sqrt{\omega}$), respectively.

These first observations provided a motivation to take a new look at the bubble theory to see how it could be re-cast in a shape which would provide more information on the common features of the two theories.

\subsection{The bubble theory: Reformulation and asymptotic approximations}
\label{subsec:asym_approx}

Work on the bubble theory begins by reformulating Eq. (\ref{eq:reaction_force}) to being an equation for the reaction force coefficient minus 1:

\begin{equation}
\label{eq:reform_reaction_force}
F-1=\frac{4(1-\tau)\beta^2}{4\tau\beta^2-9{\rm i} G} = \left( \frac{4(1-\tau)}{4\tau}\right) \left[ \frac{\beta^2}{\beta^2 - {\rm i} 9G/4\tau} \right]
\end{equation}

This structure is similar to Eqs. (\ref{eq:reform_Luong_thin}) and (\ref{eq:reform_Luong_thick}); that will be discussed in more detail in Section \ref{sec:disc}. Note the opposite sign of the complex part in the denominator; it originates from the negative sign of the complex part in the Rayleigh conductivity definition.

\begin{figure}
\centering
\includegraphics[width=6cm]{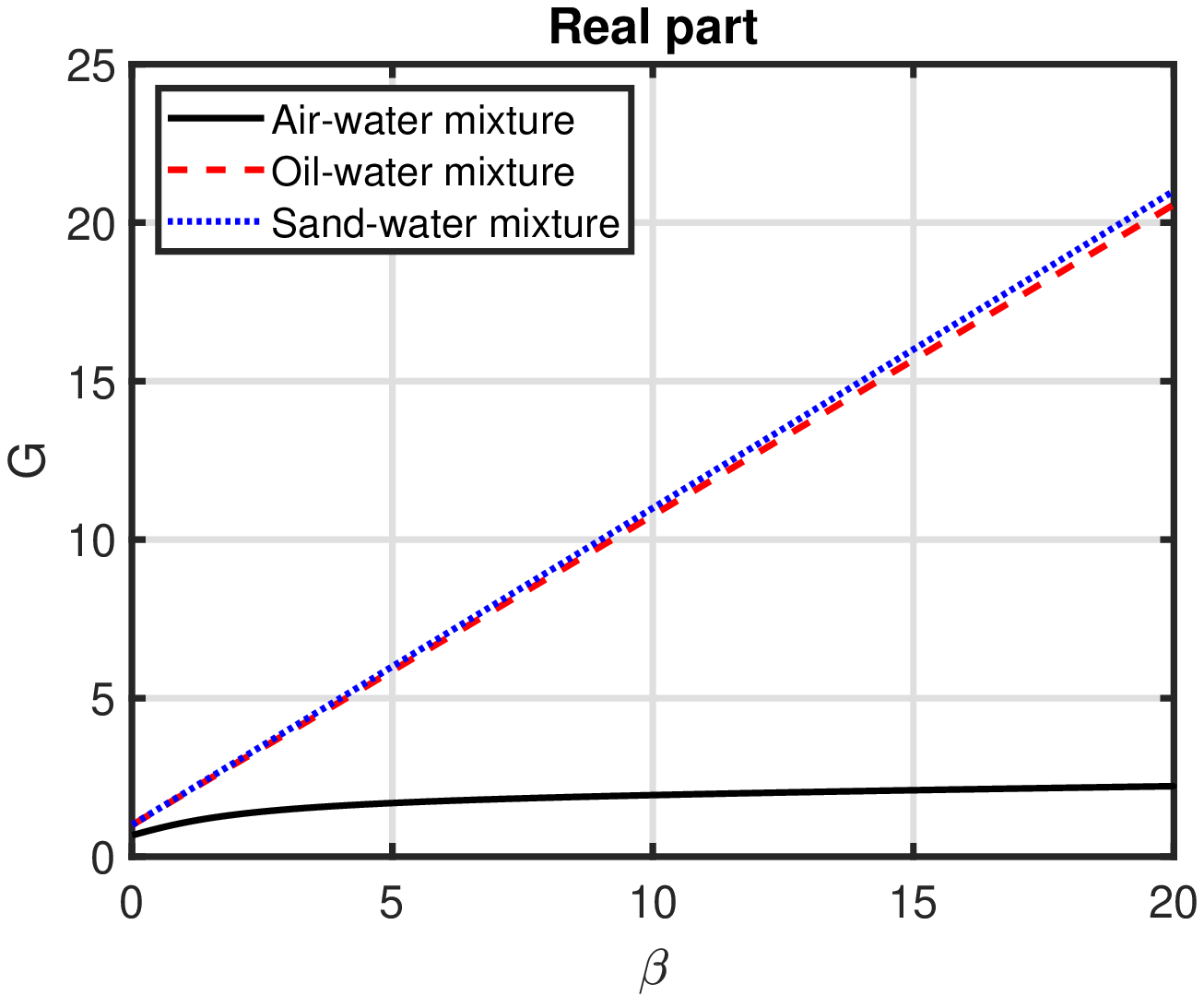}
\includegraphics[width=6cm]{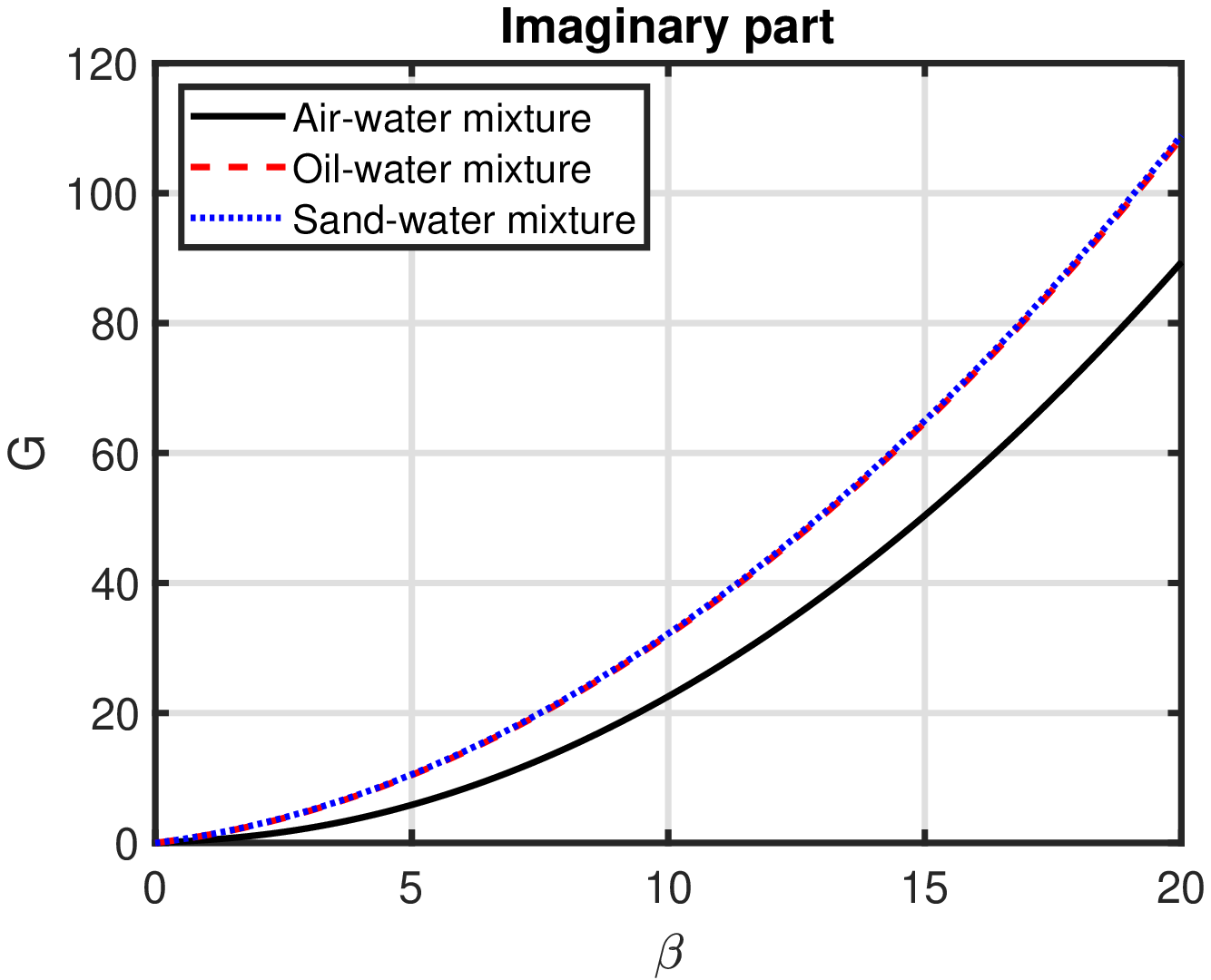}
\caption{$G$ for three mixtures, left: Real part, right: Imaginary part.}
\label{fig:G_three_mixtures}
\end{figure}

The main difference from the aperture flow is that $G$ is generally a complex number which varies with $\beta$ (and $\kappa$ for small $\kappa$), see Fig. \ref{fig:G_three_mixtures}. For $\tau=1$, i.e. equal fluid and particle density, $F=1$.

Based on Fig. \ref{fig:G_three_mixtures}, the following statements on $G$ can be made:
\begin{itemize}
  \item $G$ is a real number for $\beta = 0$
  \item Large $\beta$: Im$(G)$ $>$ Re$(G)$
  \item Re$(G)$: Small for air, larger (and almost identical) for oil and sand
  \item Im$(G)$: All three comparable, although the magnitude for air is somewhat less than for oil and sand
\end{itemize}

The real and imaginary parts of $G$ can now be written explicitly:

\begin{eqnarray}
\label{eq:g_explicit}
F-1 &=& \left( \frac{4(1-\tau)}{4\tau}\right) \left[ \frac{\beta^2}{\beta^2 + (9/4\tau){\rm Im}(G) - {\rm i} (9/4\tau) {\rm Re}(G)}  \right] \\
   &=& \frac{4(1-\tau)\beta^2}{[4\tau\beta^2 + 9{\rm Im}(G)] - {\rm i} [9 {\rm Re}(G)]} \\ \nonumber
   &=& 4(1-\tau)\beta^2 \times \frac{[4\tau\beta^2 + 9{\rm Im}(G)] + {\rm i} [9 {\rm Re}(G)]}{(4\tau\beta^2+9{\rm Im}(G))^2+(9 {\rm Re}(G))^2},
\end{eqnarray}

\noindent which can be used to write explicit equations for the real and imaginary parts of $F-1$:

\begin{eqnarray}
\label{eq:Fm1_real}
  {\rm Re}(F-1) &=& 4(1-\tau)\beta^2 \times \frac{4\tau\beta^2 + 9{\rm Im}(G)}{(4\tau\beta^2+9{\rm Im}(G))^2+(9 {\rm Re}(G))^2} \\
  \label{eq:Fm1_imag}
  {\rm Im}(F-1) &=& 4(1-\tau)\beta^2 \times \frac{9 {\rm Re}(G)}{(4\tau\beta^2+9{\rm Im}(G))^2+(9 {\rm Re}(G))^2}
\end{eqnarray}

Below, $G$ and $F-1$ will be considered for small and large $\beta$ separately; both cases can be split in two as indicated in Fig. \ref{fig:G_three_mixtures}:

\begin{itemize}
  \item Small but non-zero $\kappa$: Air
  \item Large $\kappa$: Oil and sand
\end{itemize}

\subsubsection{Small Stokes number}

A small Stokes number $\beta$ means a small $\lambda$, see Eq. (\ref{eq:lambda}). This means that the approximation that $\tanh \lambda \approx \lambda$ can be made. For $G$, terms which are either (i) constant or (ii) linear in $\lambda$ are kept, so Eq. (\ref{eq:big_g}) becomes:

\begin{eqnarray}
  G &\approx& 1 + \lambda + \frac{\lambda^2}{9} - \frac{(1+\lambda)^2}{3-2\kappa+\lambda} \\ \nonumber
   &\approx& \frac{2(\lambda+1)(1-\kappa)}{3-2\kappa+\lambda},
\end{eqnarray}

\noindent with real and imaginary parts:

\begin{eqnarray}
\label{eq:g_small_beta_real}
  {\rm Re}(G) &\approx& \frac{4(1-\kappa)\beta^2+4(1-\kappa)(2-\kappa)\beta+2(1-\kappa)(3-2\kappa)}{2\beta^2+2(3-2\kappa)\beta+(3-2\kappa)^2} \\
  \label{eq:g_small_beta_imag}
  {\rm Im}(G) &\approx& \frac{4(1-\kappa)^2\beta}{2\beta^2+2(3-2\kappa)\beta+(3-2\kappa)^2}
\end{eqnarray}

\noindent {\it Air}
\newline

For small $\kappa$, Eqs. (\ref{eq:g_small_beta_real})-(\ref{eq:g_small_beta_imag}) reduce to:

\begin{eqnarray}
\label{eq:g_small_beta_small_kappa_real}
  {\rm Re}(G) &\approx& \frac{4\beta^2+8\beta+6}{2\beta^2+6\beta+9} \approx \frac{8}{9}\beta+\frac{2}{3} \\
  \label{eq:g_small_beta_small_kappa_imag}
  {\rm Im}(G) &\approx& \frac{4\beta}{2\beta^2+6\beta+9} \approx \frac{4}{9}\beta
\end{eqnarray}

For air, $\tau \ll 1$ - combining this with Eqs. (\ref{eq:Fm1_real})-(\ref{eq:Fm1_imag}) and (\ref{eq:g_small_beta_small_kappa_real})-(\ref{eq:g_small_beta_small_kappa_imag}), results in:

\begin{eqnarray}
  {\rm Re}(F-1) &\approx& \frac{4}{9} \beta^3 \\
  {\rm Im}(F-1) &\approx& \frac{2}{3} \beta^2
\end{eqnarray}

\noindent {\it Oil and sand}
\newline

For large $\kappa$, Eqs. (\ref{eq:g_small_beta_real})-(\ref{eq:g_small_beta_imag}) reduce to:

\begin{eqnarray}
\label{eq:g_small_beta_large_kappa_real}
  {\rm Re}(G) &\approx& \frac{-4\kappa\beta^2+4\kappa^2\beta+4\kappa^2}{2\beta^2-4\kappa\beta+4\kappa^2} \approx \beta+1 \\
  \label{eq:g_small_beta_large_kappa_imag}
  {\rm Im}(G) &\approx& \frac{4\kappa^2\beta}{2\beta^2-4\kappa\beta+4\kappa^2} \approx \beta
\end{eqnarray}

For oil and sand, $\tau \sim 1$ - combining this with Eqs. (\ref{eq:Fm1_real})-(\ref{eq:Fm1_imag}) and (\ref{eq:g_small_beta_large_kappa_real})-(\ref{eq:g_small_beta_large_kappa_imag}), results in:

\begin{eqnarray}
  {\rm Re}(F-1) &\approx& \frac{4}{9} (1-\tau) \beta^3 \\
  {\rm Im}(F-1) &\approx& \frac{4}{9} (1-\tau) \beta^2
\end{eqnarray}

\noindent {\it Scaling behaviour}
\newline

Comparing the found scaling of $F-1$ with $\beta$, it is concluded that:

\begin{itemize}
  \item Scaling of real part of $K_R/2R$ and $F-1$: $(\kappa R)^2 \propto \beta^3$
  \item Scaling of imaginary part of $K_R/2R$ and $F-1$: $\kappa R \propto \beta^2$
\end{itemize}

\subsubsection{Large Stokes number}

Now the large Stokes number $\beta$ case is treated, where $\lambda$ is correspondingly large, so $\tanh \lambda \approx 1$. Now, the higher-order terms which are either (i) cubic or (ii) quartic in $\lambda$ are kept, so $G$ simplifies to:

\begin{eqnarray}
  G & \approx & 1 + \lambda + \frac{\lambda^2}{9} - \frac{\lambda-1}{1+\kappa} \\ \nonumber
   &=& \frac{2+\kappa}{1+\kappa}+\left( \frac{\kappa}{1+\kappa}\right) \lambda + \frac{\lambda^2}{9},
\end{eqnarray}

\noindent with real and imaginary parts:

\begin{eqnarray}
\label{eq:g_large_beta_real}
  {\rm Re}(G) &\approx& \left( \frac{\kappa}{1+\kappa} \right) \beta + \frac{2+\kappa}{1+\kappa} \\
  \label{eq:g_large_beta_imag}
  {\rm Im}(G) &\approx& \frac{2}{9} \beta^2 + \left( \frac{\kappa}{1+\kappa} \right) \beta
\end{eqnarray}

\noindent {\it Air}
\newline

For small $\kappa$, Eqs. (\ref{eq:g_large_beta_real})-(\ref{eq:g_large_beta_imag}) reduce to:

\begin{eqnarray}
\label{eq:g_large_beta_small_kappa_real}
  {\rm Re}(G) &\approx& \kappa \beta + 2 \\
  \label{eq:g_large_beta_small_kappa_imag}
  {\rm Im}(G) &\approx&  \frac{2}{9} \beta^2 + \kappa \beta
\end{eqnarray}

For air, $\tau \ll 1$ - combining this with Eqs. (\ref{eq:Fm1_real})-(\ref{eq:Fm1_imag}) and (\ref{eq:g_large_beta_small_kappa_real})-(\ref{eq:g_large_beta_small_kappa_imag}), results in:

\begin{eqnarray}
  {\rm Re}(F-1) &\approx& 2 \\
  {\rm Im}(F-1) &\approx& \frac{9(\kappa\beta+2)}{\beta^2}
\end{eqnarray}

\noindent {\it Oil and sand}
\newline

For large $\kappa$, Eqs. (\ref{eq:g_large_beta_real})-(\ref{eq:g_large_beta_imag}) reduce to:

\begin{eqnarray}
\label{eq:g_large_beta_large_kappa_real}
  {\rm Re}(G) &\approx&  \beta + 1 \\
  \label{eq:g_large_beta_large_kappa_imag}
  {\rm Im}(G) &\approx& \frac{2}{9} \beta^2 + \beta
\end{eqnarray}

For oil and sand, $\tau \sim 1$ - combining this with Eqs. (\ref{eq:Fm1_real})-(\ref{eq:Fm1_imag}) and (\ref{eq:g_large_beta_large_kappa_real})-(\ref{eq:g_large_beta_large_kappa_imag}), results in:

\begin{eqnarray}
  {\rm Re}(F-1) &\approx& \frac{4(1-\tau)}{4\tau+2} \\
  {\rm Im}(F-1) &\approx& \frac{36(1-\tau)}{\beta(4\tau+2)^2}
\end{eqnarray}

\noindent {\it Scaling behaviour}
\newline

Comparing the found scaling of $F-1$ with $\beta$, it is concluded that:

\begin{itemize}
  \item Scaling of real part of $K_R/2R$ and $F-1$: Both constant
  \item Scaling of imaginary part of $K_R/2R$ and $F-1$: $1/(\kappa R) \propto 1/\beta$
\end{itemize}

\subsection{Simplifications of the bubble theory}

From the approximations made in Section \ref{subsec:asym_approx}, composite expressions for the real and imaginary parts of $G$, valid for all $\beta$, can be defined.

\subsubsection{Air}

For small $\kappa$:

\begin{eqnarray}
  {\rm Re}(G) &\approx& \frac{4\beta^2+8\beta+6}{2\beta^2+6\beta+9} + \kappa \beta \\
  {\rm Im}(G) &\approx& \frac{4\beta}{2\beta^2+6\beta+9} + \frac{2}{9} \beta^2 + \kappa \beta,
\end{eqnarray}

\noindent where the further approximation $\tau \ll 1$ (air) leads to:

\begin{equation}
\label{eq:Fm1_simpl_air}
F-1 \approx \frac{4 \beta^2}{2 \beta^2+9 \left( \frac{4\beta}{2\beta^2+6\beta+9} + \kappa \beta \right) - {\rm i} 9 \left( \frac{4\beta^2+8\beta+6}{2\beta^2+6\beta+9} + \kappa \beta \right)}
\end{equation}

\subsubsection{Oil and sand}

In a similar fashion, for large $\kappa$:

\begin{eqnarray}
  {\rm Re}(G) &\approx&  \beta + 1 \\
  {\rm Im}(G) &\approx& \frac{2}{9} \beta^2 + \beta,
\end{eqnarray}

\noindent where the further assumption $\tau \sim 1$ (oil/sand) brings us to:

\begin{equation}
\label{eq:Fm1_simpl_oil_sand}
F-1 \approx \frac{4(1-\tau) \beta^2}{(4\tau+2)\beta^2 + 9\beta - {\rm i} 9 (\beta +1)}
\end{equation}

\noindent {\it F-1: Approximation versus exact expression}
\newline

\begin{figure}
\centering
\includegraphics[width=6cm]{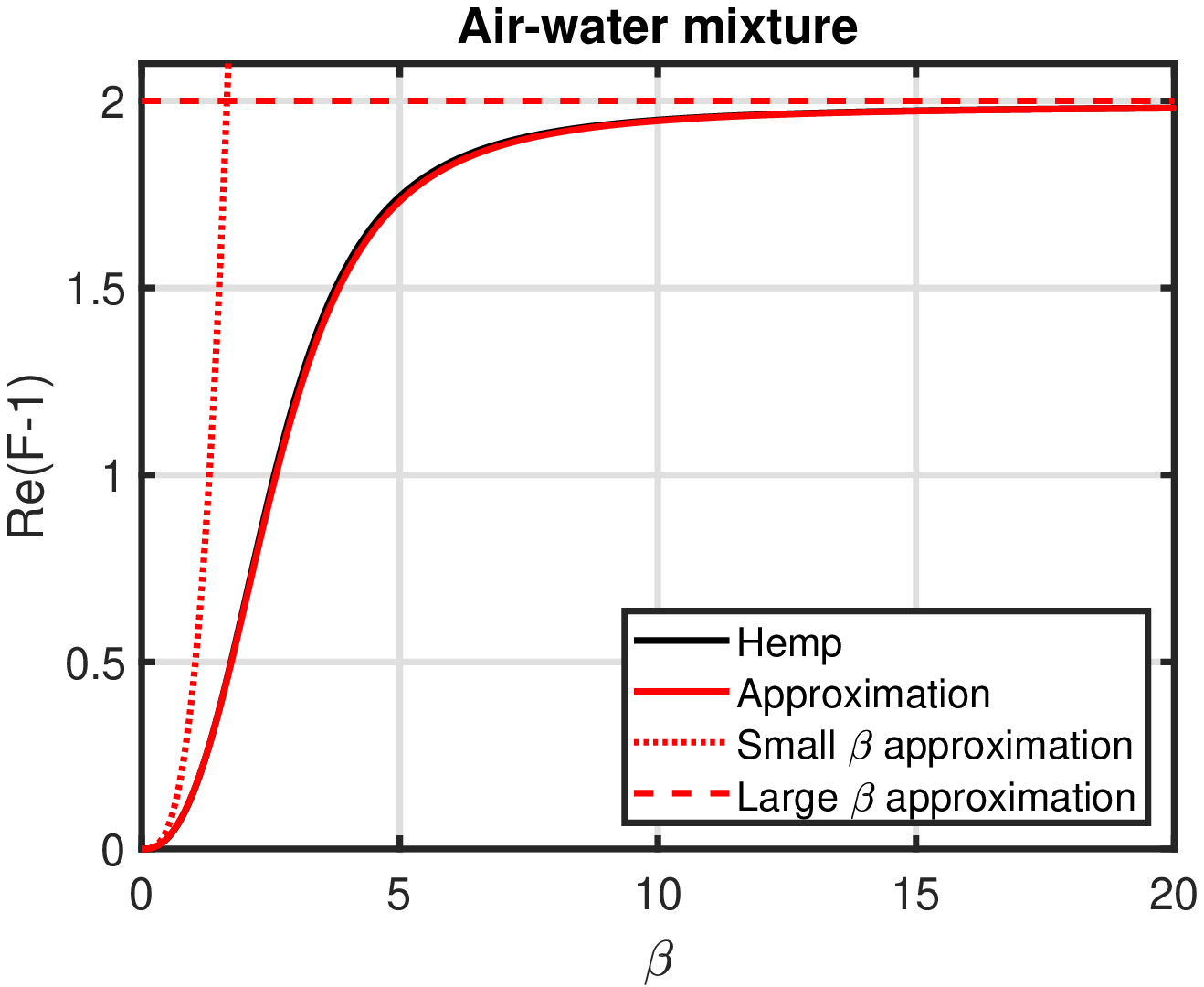}
\includegraphics[width=6cm]{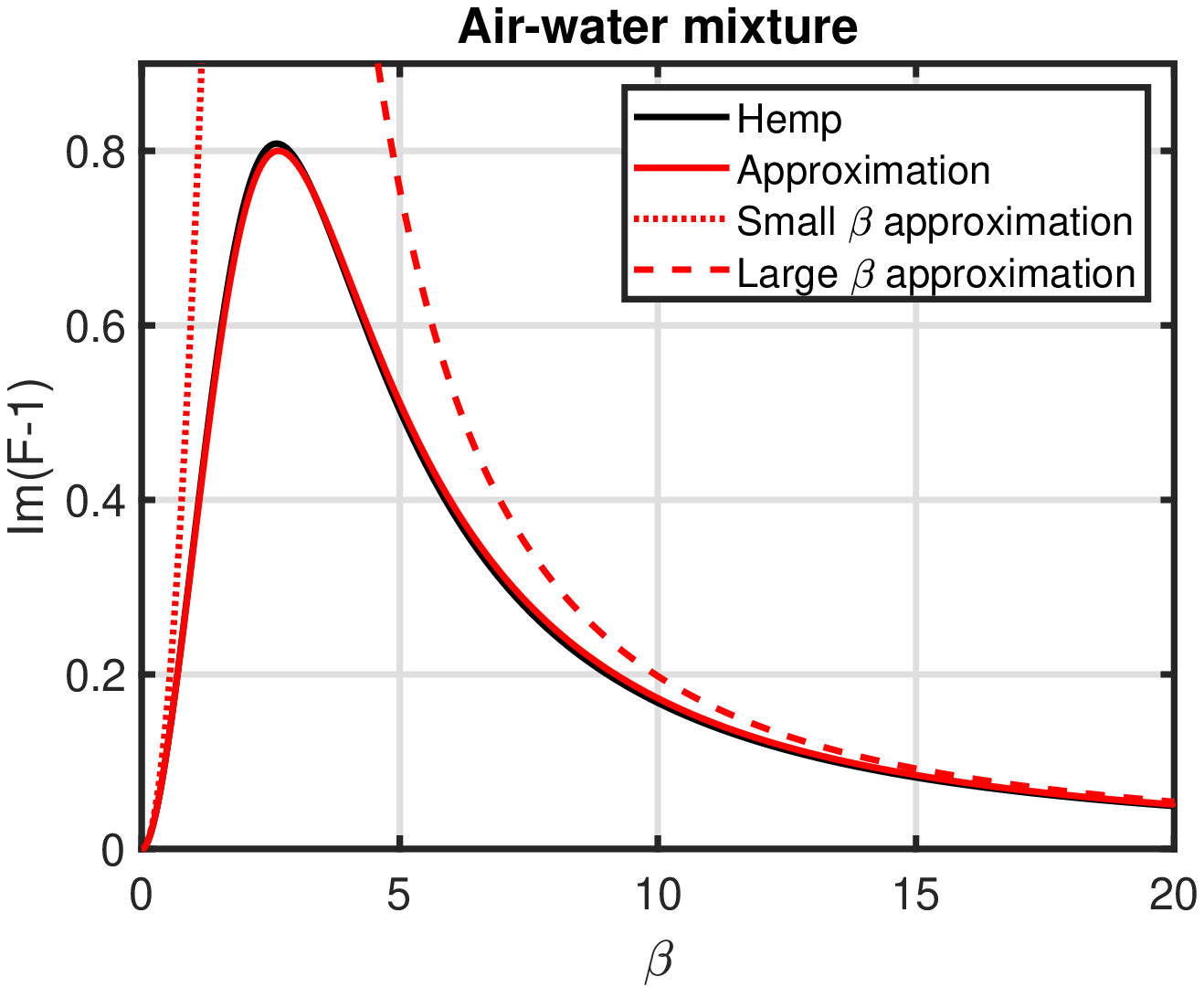}
\caption{$F-1$ for an air-water mixture, left: Real part, right: Imaginary part.}
\label{fig:Fm1_air}
\end{figure}

\begin{figure}
\centering
\includegraphics[width=6cm]{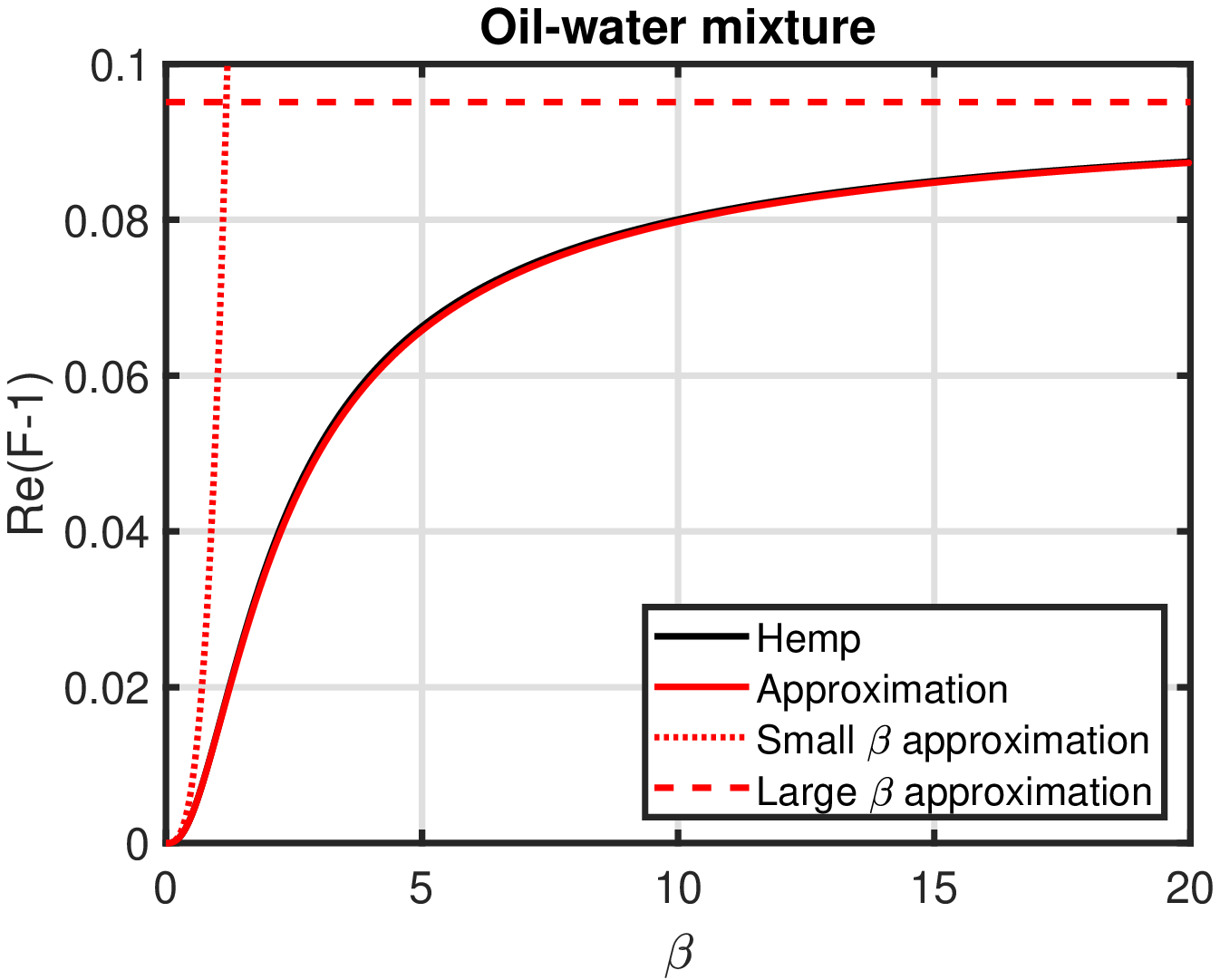}
\includegraphics[width=6cm]{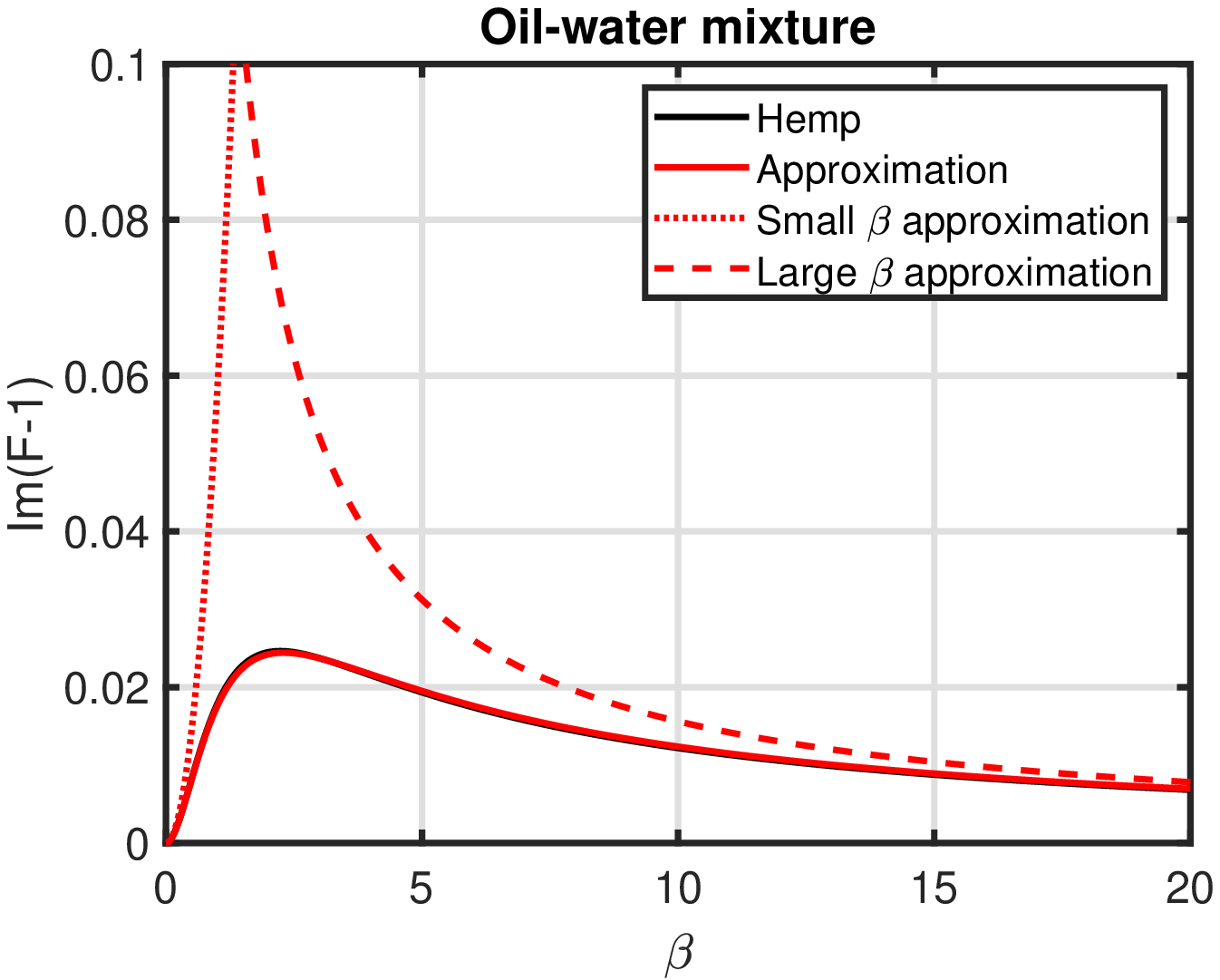}
\caption{$F-1$ for an oil-water mixture, left: Real part, right: Imaginary part.}
\label{fig:Fm1_oil}
\end{figure}

\begin{figure}
\centering
\includegraphics[width=6cm]{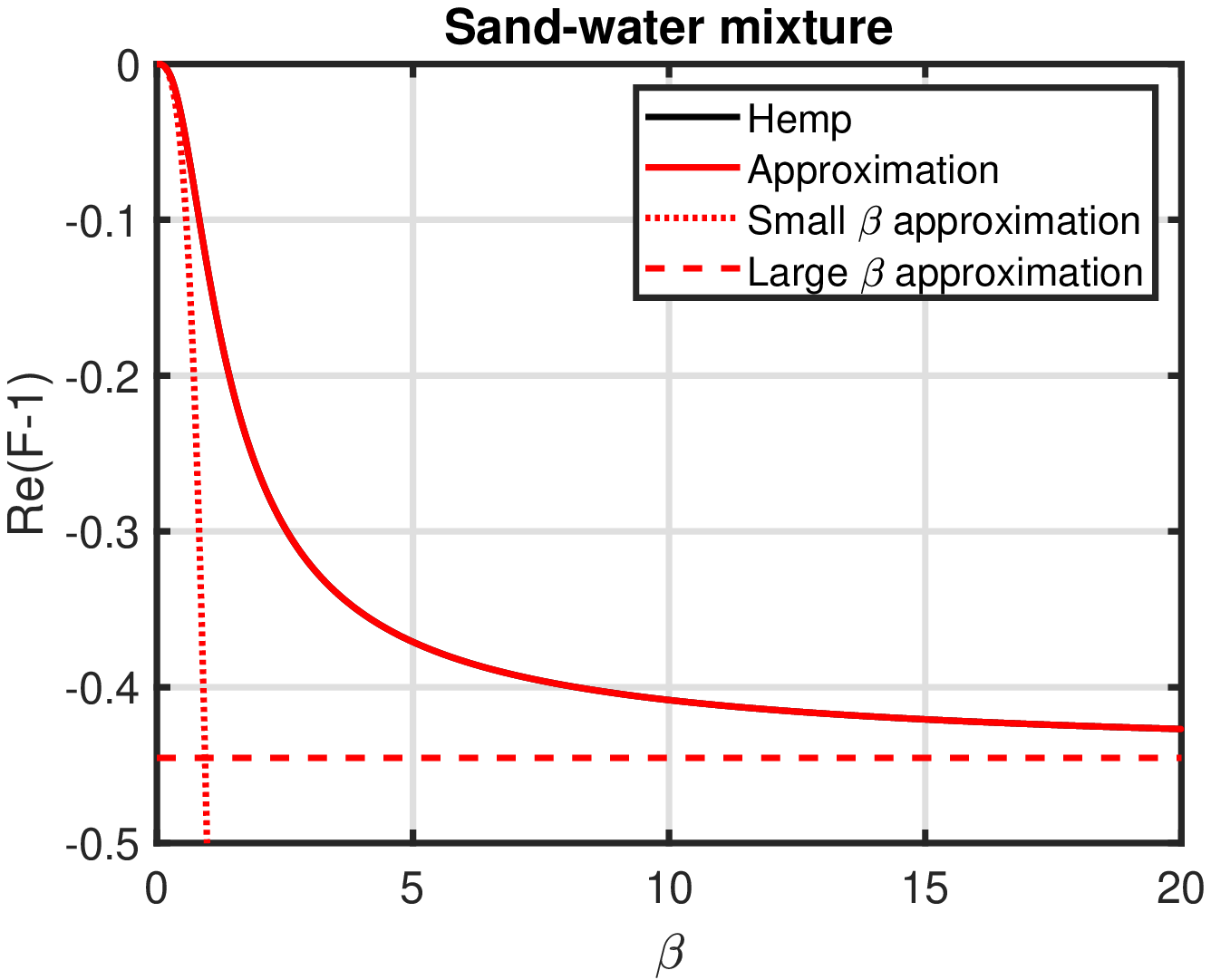}
\includegraphics[width=6cm]{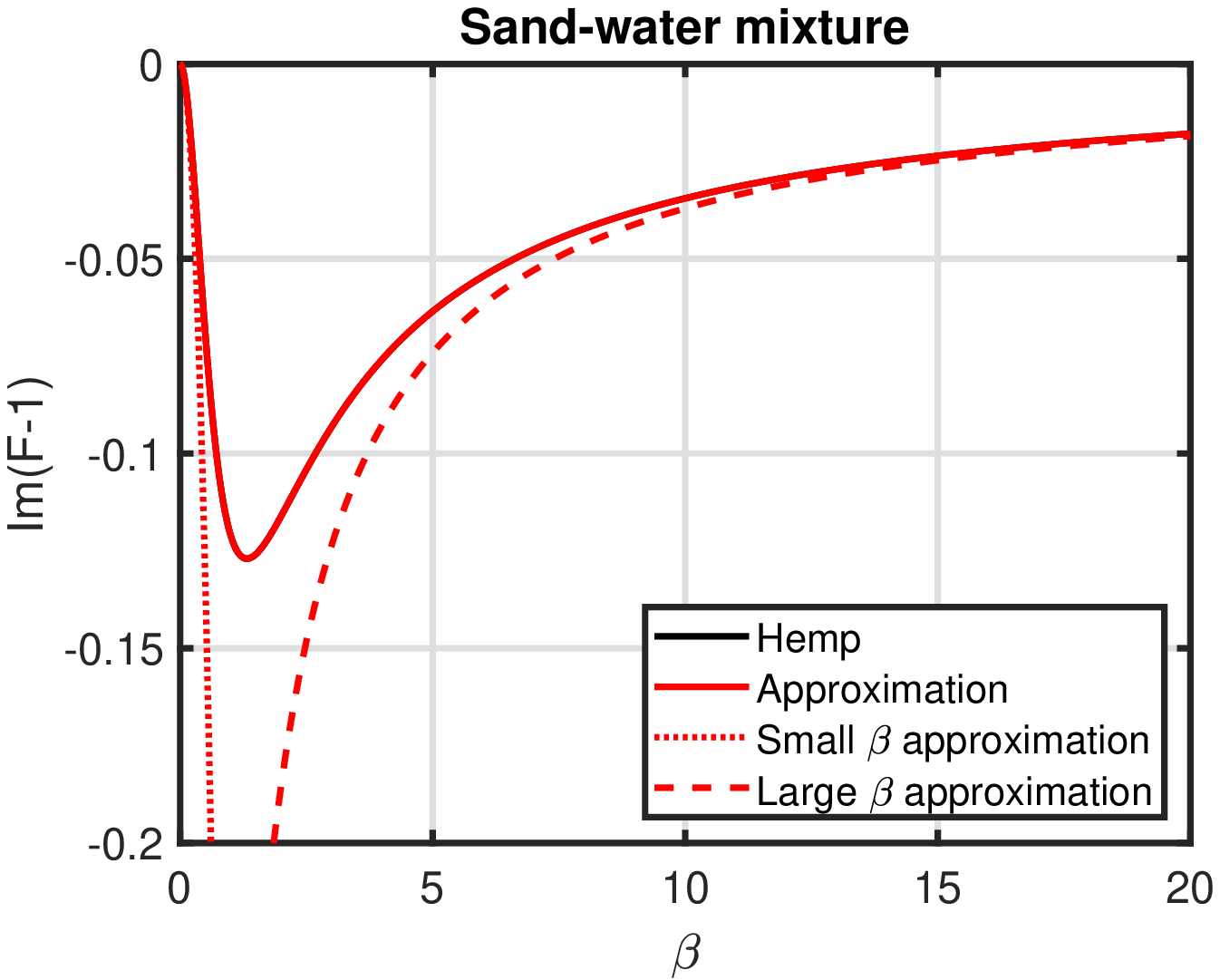}
\caption{$F-1$ for a sand-water mixture, left: Real part, right: Imaginary part.}
\label{fig:Fm1_sand}
\end{figure}

The approximations of $F-1$ from Eqs. (\ref{eq:Fm1_simpl_air}) and (\ref{eq:Fm1_simpl_oil_sand}) compared to the exact $F-1$ are plotted in Figs. \ref{fig:Fm1_air}-\ref{fig:Fm1_sand}: The approximations are very close to the exact values. Also shown are the asymptotic approximations from Section \ref{subsec:asym_approx}. Note that the asymptotic approximations for small $\beta$ are valid for $\beta < 0.1$.

\section{Discussion}
\label{sec:disc}

\subsection{Comparison of the two phenomena}
\label{subsec:comp_phen}

\subsubsection{Structure}

As mentioned at the beginning of Section \ref{subsec:asym_approx}, the structure of the normalised Rayleigh conductivity and the reaction force coefficient (minus 1) is similar. This is summarised in Table \ref{tab:luong_hemp_list}. From the table it is observed that the Strouhal number $\kappa R$ is proportional to the squared Stokes number $\beta^2$, but from the asymptotic formulae these scalings have been found:

\begin{itemize}
  \item Small $\beta$, real part: $\kappa R \propto \beta^{3/2}$
  \item Small $\beta$, imaginary part: $\kappa R \propto \beta^2$
  \item Large $\beta$, real part: No scaling with $\kappa R$ and $\beta$
  \item Large $\beta$, imaginary part: $\kappa R \propto \beta$
\end{itemize}

The scaling $\kappa R \propto \beta$ is more likely if the length scales $R$ and $a$ are dominating, while the scaling $\kappa R \propto \beta^2$ matches the angular frequency $\omega$, see Eqs. (\ref{eq:Strouhal}) and (\ref{eq:Stokes}).

\begin{table}
\caption{Luong et al. and Hemp comparison.}
\centering
\begin{tabular}{cc}
\hline
\textbf{Luong et al. (thin wall)}	& \textbf{Hemp}\\
\hline
$K_R/2R$    &   $F-1$\\
\\
 $\omega R/U$   &   $\beta^2$\\
\\
 1 &   $ 4(1-\tau)/4\tau$\\
\\
 $2/\pi \sigma^2$  & $-9G/4\tau$\\
\hline
\end{tabular}
\label{tab:luong_hemp_list}
\end{table}

The $F-1$ structure is equal to the normalised Rayleigh conductivity if $G$ is a real number. This is not the case in general, but it does occur for $\beta=0$, see Eqs. (\ref{eq:g_small_beta_small_kappa_real}) and (\ref{eq:g_small_beta_large_kappa_real}).

Generalising Stokes law (Eq. (\ref{eq:drag_Stokes})) for low Reynolds number to include viscosity in the particle \cite{howe_b}:

\begin{equation}
F_{D,~{\rm low~Re}}=-u_p(4[C] \pi  \mu_f a) = -u_p \left (4 \left[ \frac{2 \mu_f + 3 \mu_p}{2(\mu_f + \mu_p)} \right] \pi \mu_f a \right)
\end{equation}

\noindent {\it Air}
\newline

From Eq. (\ref{eq:g_small_beta_small_kappa_real}), $G=2/3$ for $\beta=0$. If $\kappa = \mu_p/\mu_f \ll 1$:

\begin{equation}
C=\frac{2 \mu_f + 3 \mu_p}{2(\mu_f + \mu_p)}=1,
\end{equation}

\noindent leading to:

\begin{equation}
F_{D,~{\rm low~Re}}=-u_p(4 \pi  \mu_f a)
\end{equation}

\noindent {\it Oil and sand}
\newline

From Eq. (\ref{eq:g_small_beta_large_kappa_real}), $G=1$ for $\beta=0$. If $\kappa = \mu_p/\mu_f \gg 1$:

\begin{equation}
C=\frac{2 \mu_f + 3 \mu_p}{2(\mu_f + \mu_p)}=3/2,
\end{equation}

\noindent which leads to the Stokes equation:

\begin{equation}
F_{D,~{\rm low~Re}}=-u_p(6 \pi  \mu_f a)
\end{equation}

Thus, for $F-1$ to match the normalised Rayleigh conductivity, the drag term has to be equal (or proportional) to the Stokes drag.

\subsubsection{Physical picture}

The two physical phenomena are discussed, and how the characteristic quantities can be seen as corresponding to each other.

It should be kept in mind that there are significant differences as well, e.g that the bias flow aperture theory is for a single-phase high Reynolds number flow and the bubble theory is for a two-phase flow at low Reynolds numbers. However, one could think of the bias aperture flow as also consisting of two "phases", where one is the mean flow and the other is the vortices.

For the aperture theory, vortex shedding leads to blockage and acoustic damping - for the bubble theory, drag induced by the oscillating container leads to decoupled motion of particles and fluid, which in turn leads to measurement errors and damping.

For bias aperture flow, an oscillating pressure across the aperture leads to vortex shedding from the aperture rim with the same frequency. For the bubble theory, the oscillation of the container leads to the decoupled motion of the particles from the fluid.

In both cases, a mean flow has to exist for the physical effect to take place; for the aperture theory, the mean flow is parallel to the oscillation, whereas the flow for the bubble theory is perpendicular to the oscillation.

The characteristic scale for the aperture is the radius, for the bubble theory it is the particle radius.

For aperture flow, a jet is formed downstream, which encompasses the vorticity and sweeps it away. One might speculate that the analogy for the bubble theory may be the wakes of the particles when they execute their decoupled motion. So the contraction ratio (Eq. (\ref{eq:contraction_ratio})) may have an equivalent "wake ratio" for the bubble theory:

\begin{equation}
\Sigma = \frac{\mathcal{W}}{\pi a^2},
\end{equation}

\noindent where $\mathcal{W}$ is the minimum wake area. However, since the bubble theory is derived for low Reynolds number, the wake picture may not be precise. The particles may interact with their own wakes which presents an additional complication.

\subsubsection{How to match the bubble theory to the aperture theory}

A small exercise to determine the bubble theory parameters needed to match $F-1$ to the normalised Rayleigh conductivity (Eq. (\ref{eq:reform_Luong_thin})) is carried out.

From Table \ref{tab:luong_hemp_list}, first the required $\tau$ is found:

\begin{eqnarray}
1 &=& \frac{4(1-\tau)}{4\tau} \\
  \tau &=& 1/2
\end{eqnarray}

This means that the density of the particle is half of the density of the fluid. Using this $\tau$, $G$ can be expressed using $\sigma$:

\begin{eqnarray}
\label{eq:g_exercise_one}
  \frac{2}{\pi \sigma^2} &=& \frac{9G}{4\tau} \\
  \label{eq:g_exercise_two}
  G &=& \frac{8 \tau}{9 \pi \sigma^2}
\end{eqnarray}

Using $\tau=1/2$ in Eq. (\ref{eq:g_exercise_two}), $G=0.25$, which corresponds to one-fourth of the Stokes drag, see Eq. (\ref{eq:drag_Stokes}). This value of $G$ is outside the standard range, which is between $2/3$ ($\kappa \ll 1$) and $1$ ($\kappa \gg 1$). Another point of view would be to consider whether the contraction ratio $\sigma$ might be smaller than 0.75. Setting $G$ to $2/3$ ($1$), the corresponding $\sigma$ is 0.46 (0.38), respectively. Recent simulations of a laminar viscous jet through an aperture do indeed show that $\sigma$ decreases with decreasing Reynolds number \cite{fabre_a}: For $\sigma \approx 0.4$, $Re \approx 10$.

To conclude, the bubble theory matches the bias-flow aperture theory if:

\begin{itemize}
\item $\beta \approx 0$
  \item The particle density is half of the fluid density
  \item (i) $G$ is a real number equal to $0.25$, i.e. one-fourth of the Stokes drag or (ii) the contraction ratio $\sigma \approx 0.4$
\end{itemize}

\subsection{Further simplifications of the bubble theory approximations}
\label{subsec:further_simpl}

The approximations of $F-1$ are further simplified to arrive at expressions with a constant imaginary part in denominator.

\subsubsection{Air}

Assuming large $\beta$, Eq. (\ref{eq:Fm1_simpl_air}) is simplified to:

\begin{eqnarray}
  F-1 &\approx& \frac{4 \beta^2}{2 \beta^2 - {\rm i} 9 (2+\kappa \beta)} \\ \nonumber
   &\approx& \frac{2 \beta^2}{\beta^2 - {\rm i} 9},
\end{eqnarray}

\noindent which means that $\beta$ for maximum damping is:

\begin{equation}
\beta_{\rm max~damping} = 3,
\end{equation}

\noindent see Table \ref{tab:hemp_exact_approx}. Then the real and imaginary parts are written explicitly:

\begin{eqnarray}
\label{eq:real_air_further_simpl}
  {\rm Re}(F-1) &\approx& \frac{2\beta^4}{\beta^4+9^2} \\
  \label{eq:imag_air_further_simpl}
  {\rm Im}(F-1) &\approx& \frac{18\beta^2}{\beta^4+9^2}
\end{eqnarray}

\subsubsection{Oil and sand}

Assuming large $\beta$, Eq. (\ref{eq:Fm1_simpl_oil_sand}) is simplified to:

\begin{eqnarray}
  F-1 &\approx& \frac{4(1-\tau)\beta^2}{(4\tau+2)\beta^2-{\rm i} 9 (\beta+1)} \\ \nonumber
   &\approx& \frac{4(1-\tau)}{4\tau+2} \left[ \frac{\beta^2}{\beta^2-{\rm i} 9/(4\tau+2)} \right],
\end{eqnarray}

\noindent which means that $\beta$ for maximum damping is:

\begin{equation}
\beta_{\rm max~damping} = \sqrt{\frac{9}{4\tau+2}},
\end{equation}

\noindent see Table \ref{tab:hemp_exact_approx}. Then the real and imaginary parts are written explicitly:

\begin{eqnarray}
\label{eq:real_oil_sand_further_simpl}
  {\rm Re}(F-1) &\approx& \frac{4(1-\tau)}{4\tau+2} \left[ \frac{\beta^4}{\beta^4+\left( \frac{9}{4 \tau +2} \right)^2} \right] \\
  \label{eq:imag_oil_sand_further_simpl}
  {\rm Im}(F-1) &\approx& \frac{36(1-\tau)}{(4\tau+2)^2} \left[ \frac{\beta^2}{\beta^4+\left( \frac{9}{4 \tau +2} \right)^2} \right]
\end{eqnarray}

\begin{table}
\caption{Exact and simplified $\beta_{\rm max~damping}$.}
\centering
\begin{tabular}{ccc}
\hline
\textbf{Particle}	& \textbf{Hemp $\beta_{\rm max~damping}$ (exact)} & \textbf{Hemp $\beta_{\rm max~damping}$ (simplified)}\\
\hline
Air    & 2.6 & 3.0 \\
\\
Oil    & 2.2 & 1.3 \\
\\
Sand   & 1.3 & 0.9 \\
\hline
\end{tabular}
\label{tab:hemp_exact_approx}
\end{table}

\begin{figure}
\centering
\includegraphics[width=6cm]{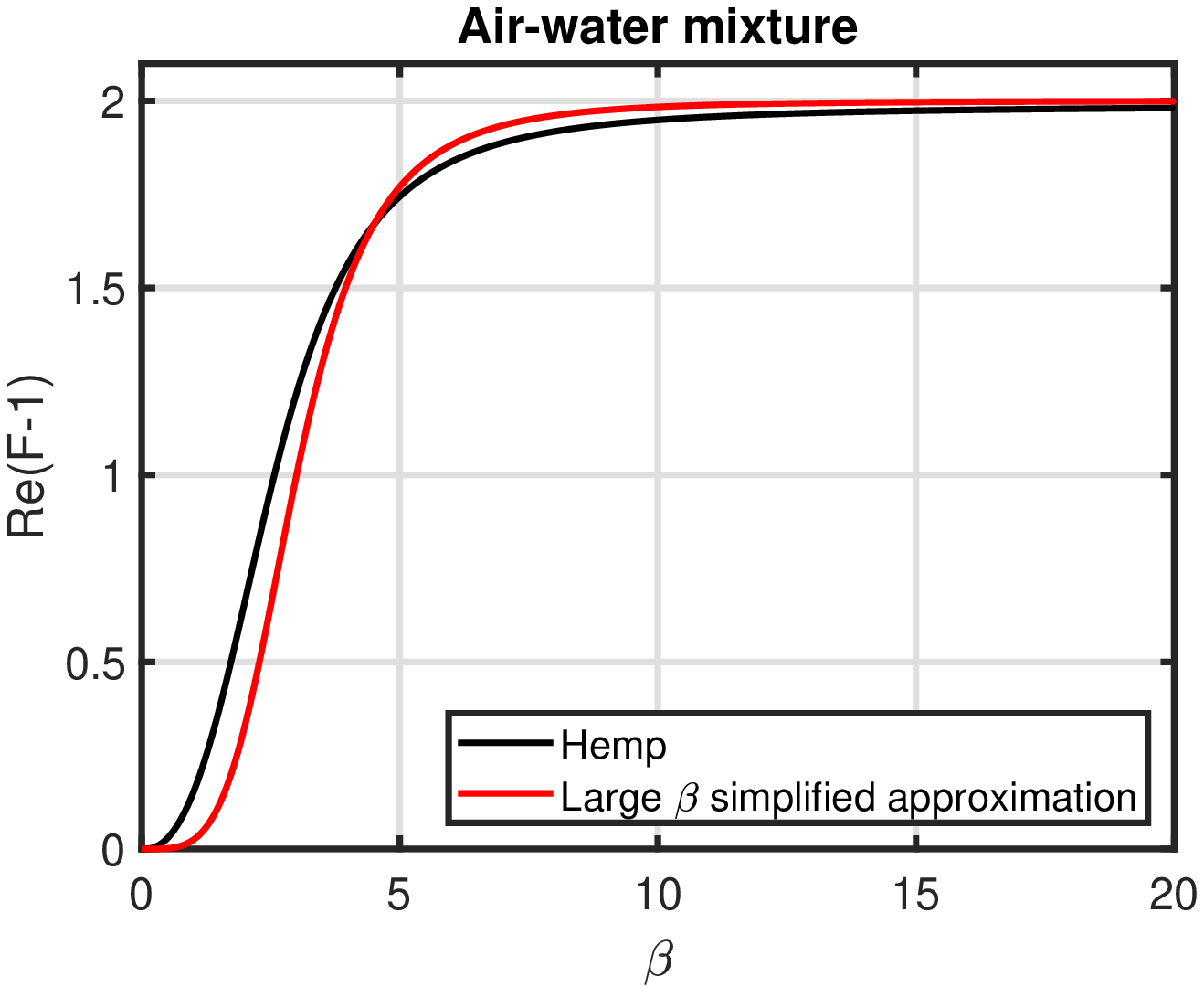}
\includegraphics[width=6cm]{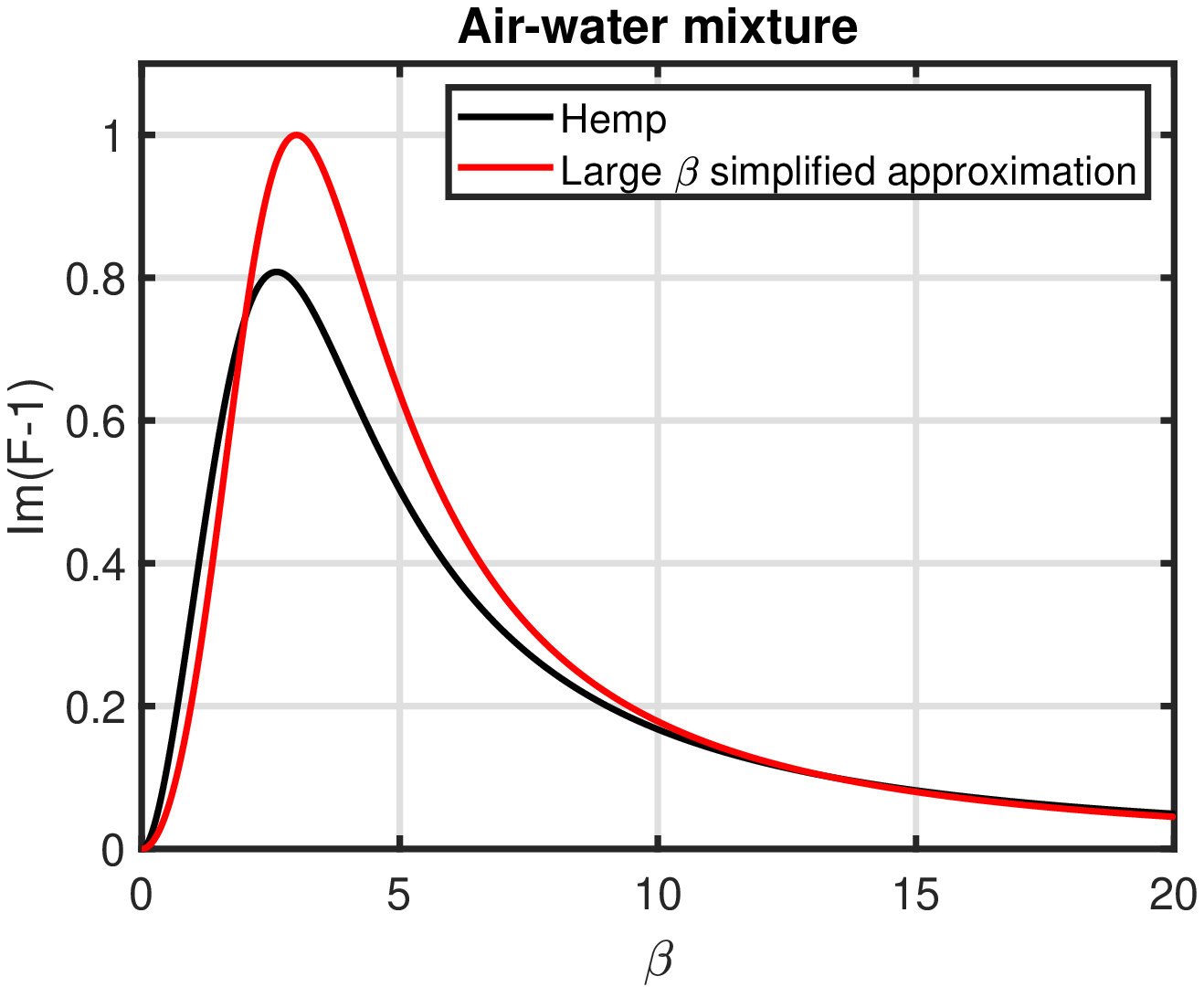}
\caption{$F-1$ for an air-water mixture, left: Real part, right: Imaginary part.}
\label{fig:simpl_Fm1_air}
\end{figure}

\begin{figure}
\centering
\includegraphics[width=6cm]{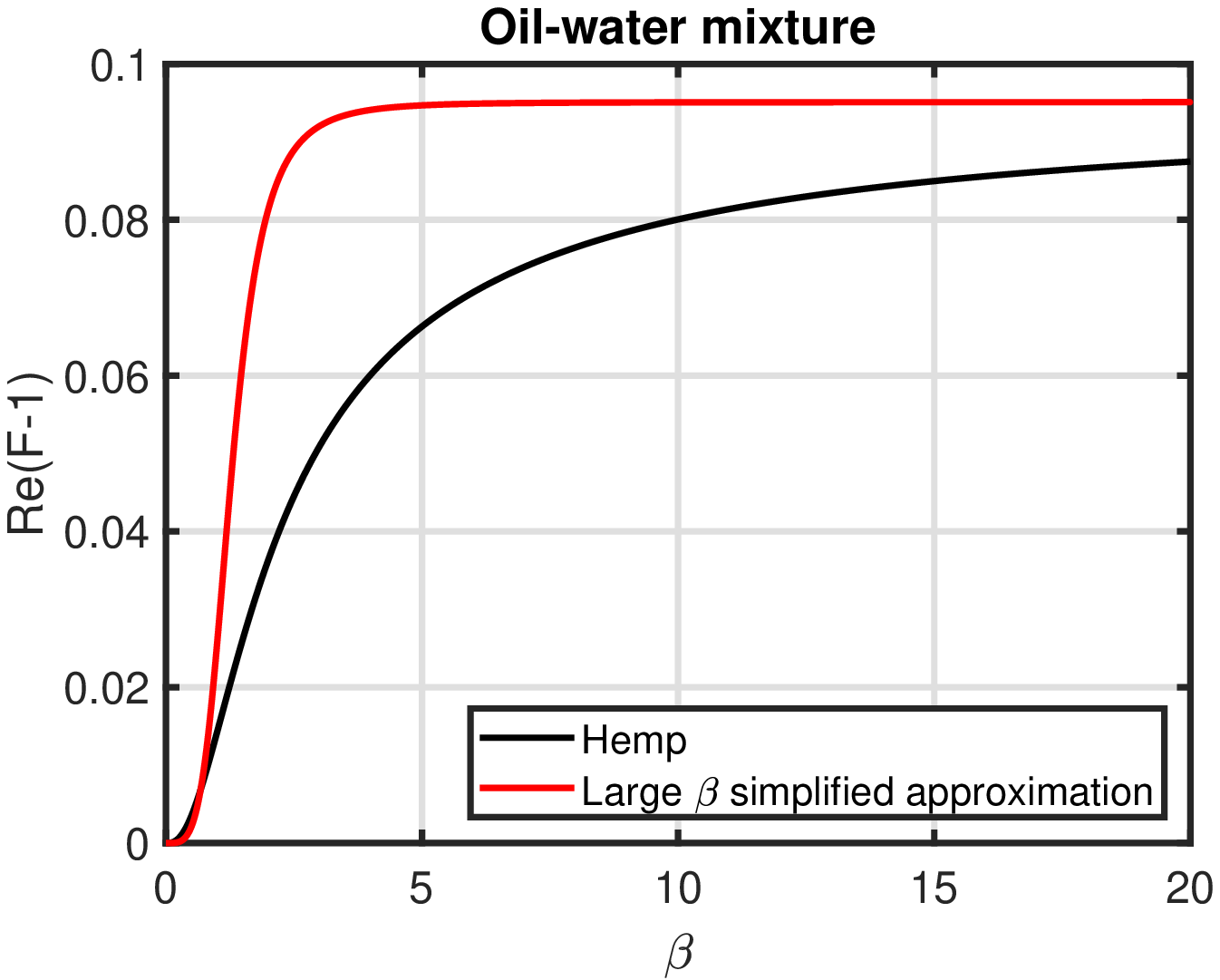}
\includegraphics[width=6cm]{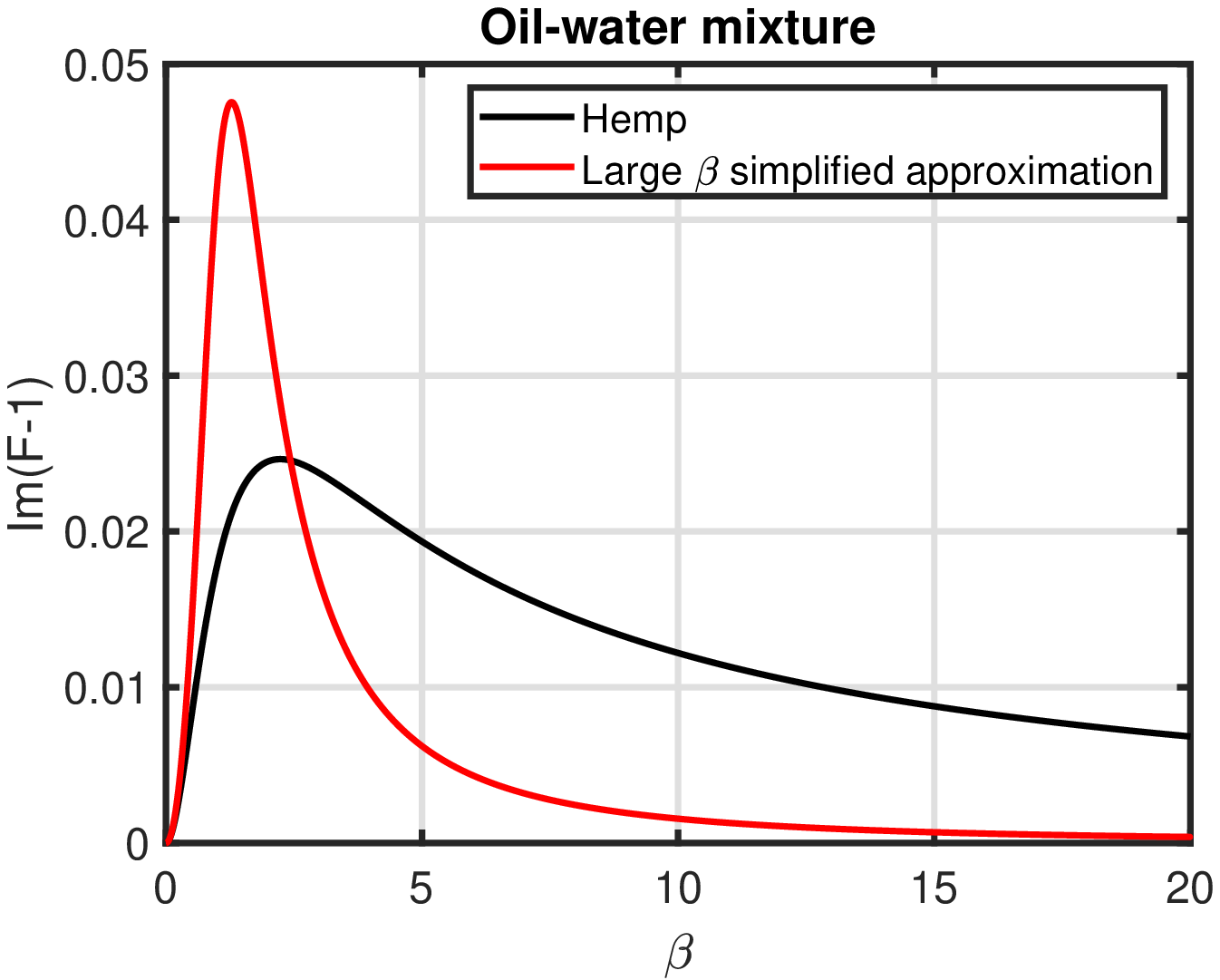}
\caption{$F-1$ for an oil-water mixture, left: Real part, right: Imaginary part.}
\label{fig:simpl_Fm1_oil}
\end{figure}

\begin{figure}
\centering
\includegraphics[width=6cm]{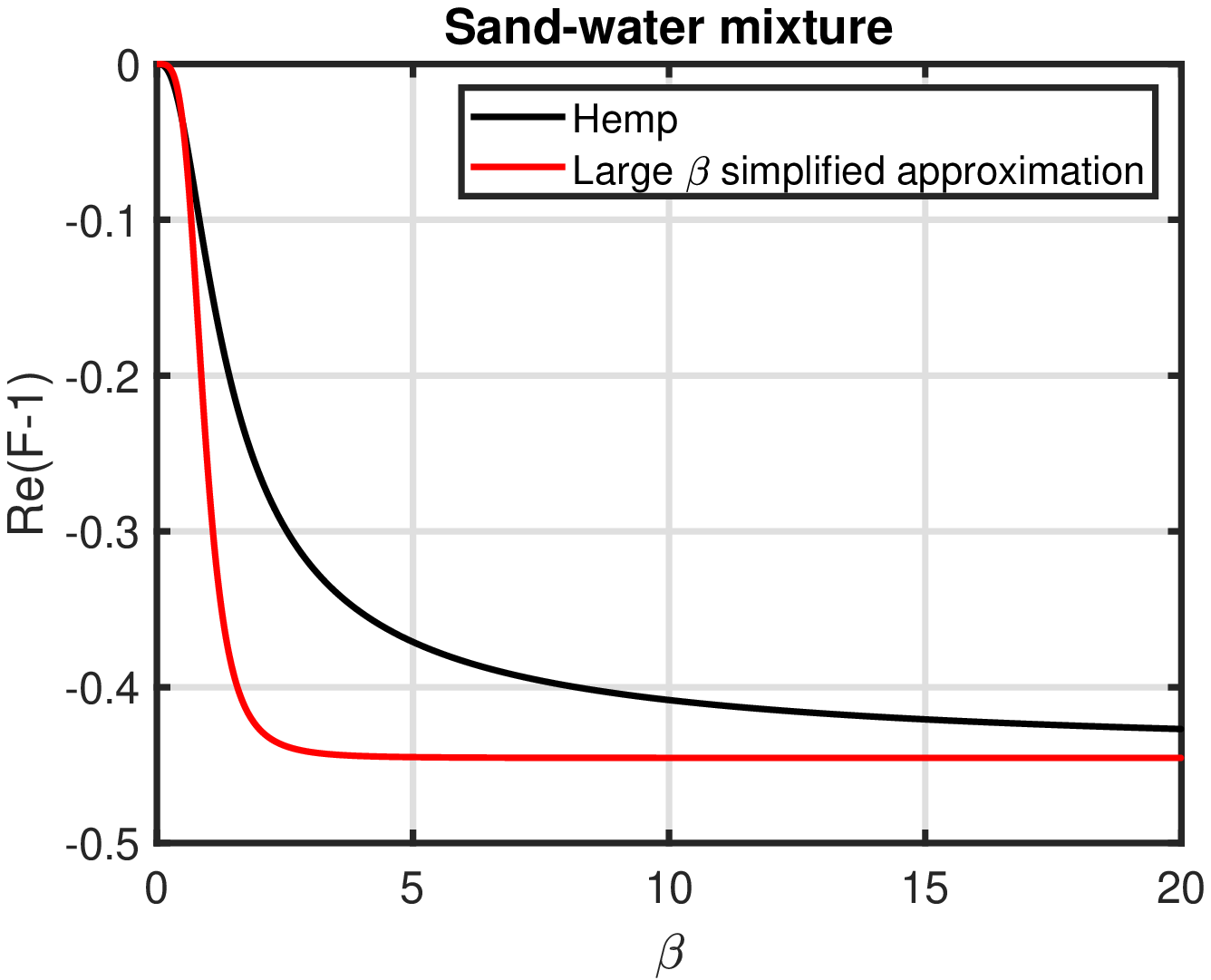}
\includegraphics[width=6cm]{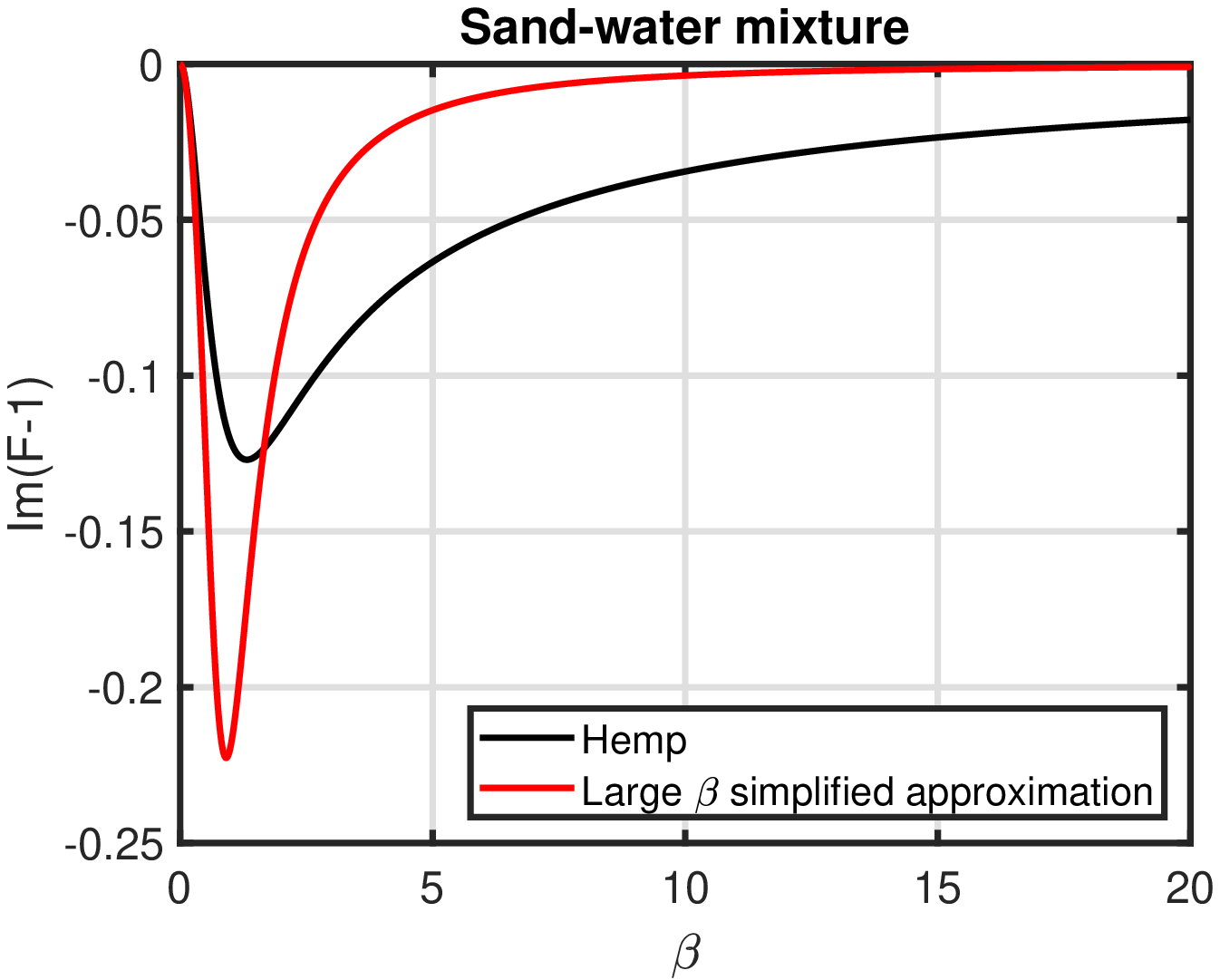}
\caption{$F-1$ for a sand-water mixture, left: Real part, right: Imaginary part.}
\label{fig:simpl_Fm1_sand}
\end{figure}

The simplified approximations are compared to the exact cases in Figs. \ref{fig:simpl_Fm1_air} - \ref{fig:simpl_Fm1_sand}: The best agreement is found for the air-water mixture.

\subsection{Applications to Coriolis flowmetering}

The simplified approximations in Section \ref{subsec:further_simpl} can be used to create simple Stokes-number-dependent expressions for Coriolis flowmeter measurement errors and damping due to two-phase flow.

\subsubsection{Measurement error}

The mass flow ($E_{\dot{m}}$) and density ($E_d$) error due to entrained particles is proportional to the real part of $1-F$ \cite{basse_a}:

\begin{equation}
E_{\dot{m}}=E_d=\frac{\alpha (\rho_f-\rho_p) {\rm Re}(1-F)}{\alpha \rho_p + (1-\alpha)\rho_f},
\end{equation}

\noindent where $\alpha$ is the volumetric particle fraction.

\subsubsection{Damping}

Damping due to entrained particles is proportional to the imaginary part of $F$ \cite{basse_b} - the work done per cycle on the particles by the container is:

\begin{equation}
 W_p = \pi (\rho_f-\rho_p) \alpha V_{f-p} \omega^2 u^2 {\rm Im}(F),
\end{equation}

\noindent where $V_{f-p}$ is the volume of the fluid-particle ($f-p$) mixture and $u$ is the amplitude of the container oscillation.

\subsubsection{Air}

For an air-water mixture, the measurement error and damping are given by the following two equations:

\begin{equation}
E_{\dot{m}}=E_d \approx - \frac{\alpha (\rho_f-\rho_p)}{\alpha \rho_p + (1-\alpha)\rho_f} \left[ \frac{2\beta^4}{\beta^4+9^2} \right]
\end{equation}

\begin{equation}
 W_p \approx \pi (\rho_f-\rho_p) \alpha V_{f-p} \omega^2 u^2 \left[ \frac{18\beta^2}{\beta^4+9^2} \right]
\end{equation}

\subsubsection{Oil and sand}

For an oil-water or sand-water mixture, the measurement error and damping are given by the following two equations:

\begin{equation}
E_{\dot{m}}=E_d \approx - \frac{\alpha (\rho_f-\rho_p)}{\alpha \rho_p + (1-\alpha)\rho_f} \left[ \frac{4(1-\tau)}{4\tau+2} \left( \frac{\beta^4}{\beta^4+\left( \frac{9}{4 \tau +2} \right)^2} \right) \right]
\end{equation}

\begin{equation}
 W_p \approx \pi (\rho_f-\rho_p) \alpha V_{f-p} \omega^2 u^2 \left[ \frac{36(1-\tau)}{(4\tau+2)^2} \left( \frac{\beta^2}{\beta^4+\left( \frac{9}{4 \tau +2} \right)^2} \right) \right]
\end{equation}

\section{Conclusions}
\label{sec:conc}

The analogy between two theories, the bias flow aperture theory and the Coriolis flowmeter "bubble theory", has been explored. Both theories are developed for incompressible flow, but the aperture theory is for single-phase, high Reynolds number flow, whereas the bubble theory is valid for two-phase, low Reynolds number flow.

The aperture theory deals with oscillating pressure generating vortex shedding, which acts to block the flow and dampen sound; the bubble theory shows how particle drag in an oscillating fluid leads to decoupled particle motion, which in turn leads to measurement errors and damping in Coriolis flowmeters.

The bubble theory has been simplified to allow a more direct comparison to the aperture theory. The comparison is summarised in Section \ref{subsec:comp_phen}. An example illustrates which conditions are necessary for the two theories to match. There are indications that low Reynolds number bias flow aperture simulations \cite{fabre_a} correspond closer to the bubble theory than the high Reynolds number bias flow aperture theory.

Simplified expressions for the bubble theory have been derived in analogy with the simplifications of the aperture theory presented in \cite{luong_a}.

\section*{Acknowledgements}

The author is grateful to Dr. John Hemp for creating, providing and explaining/discussing the Coriolis flowmeter bubble theory \cite{hemp_a}. The author also appreciates the patience of his family during the process of writing this paper.


\end{document}